\documentclass[twocolumn,
               aps,
               prd,   
               preprintnumbers,
               superscriptaddress,
               numbers,
               floatfix,
               nofootinbib,
               showpacs,
               colorlinks,
               linkcolor=blue,   
               citecolor=blue]{revtex4-2}

\usepackage{graphicx}
\graphicspath{{Figures/}}

\usepackage{wrapfig}
\usepackage{float}
\usepackage[utf8x]{inputenc}
\usepackage{amsmath}
\usepackage{orcidlink}
\usepackage{amssymb}
\usepackage{amsthm}
\usepackage{mathtools}
\usepackage{physics}
\usepackage{color}
\usepackage{xcolor}
\usepackage{multirow}
\usepackage{physics}
\usepackage{tabularx}
\usepackage{comment}
\usepackage{soul}
\usepackage{siunitx}
\usepackage{threeparttable}

\DeclareSIUnit\bar{bar}
\DeclareSIUnit{\year}{year}

\begin{document}

\title{Broadband interferometry-based searches for photon-axion conversion in vacuum}

\author{Josep Maria Batllori\orcidlink{0000-0003-0555-8970}}
\email{Corresponding author: josep.batllori.berenguer@uni-hamburg.de}

\author{Dieter Horns\orcidlink{0000-0003-1945-0119}}

\author{Marios Maroudas\orcidlink{0000-0003-1294-1433}}

\affiliation{Institut für Experimentalphysik, Universität Hamburg, Luruper Chaussee 149, D-22761 Hamburg, Germany}

\date{\today}

\begin{abstract}
A novel experiment is introduced to detect photon-axion conversion independent of the dark-matter hypothesis in a broad mass-range called WISP Interferometer (WINTER). The setup consists of a free-space Mach-Zehnder-type interferometer incorporating an external magnetic field and vacuum in one of the arms, where photon-axion mixing occurs via the Primakoff effect and is detected through changes in amplitude. The expected axion-induced signal is then modulated by polarization changes. The experiment is designed to integrate a Fabry-P\'erot cavity with a finesse of $10^{5}$ that will be operated in a vacuum environment, significantly enhancing the sensitivity. It is projected to reach the Dine-Fischler-Srednicki-Zhitnitsky theoretical line with photon-axion coupling sensitivities down to $g_{a\gamma\gamma}\simeq 3.7\times10^{-14}~\si{\GeV^{-1}}$ for axion masses up to \SI{380}{\micro \eV}.
\end{abstract}


\maketitle

\section{Introduction}
\label{sec:introduction}

Axions are hypothetical particles that arise in various extensions of the Standard Model of particle physics. Originally proposed to resolve the strong CP problem in QCD \cite{peccei_mathrmcp_1977, kim_weak-interaction_1979}, axions have become a prominent candidate for dark matter due to their weak interactions with normal matter and their ability to exist in large quantities in the Universe without being directly detected \cite{dine_simple_1981, Sikivie_experimental_1983}. A key aspect of detecting axions is their coupling to photons in the presence of an external magnetic field. This interaction allows for the conversion of photons into axions and vice versa when they traverse regions with strong magnetic fields.

Several experimental approaches have been developed to exploit this axion-photon coupling. Haloscopes aim to detect dark matter axions by converting them into photons using a low-loss resonant cavity embedded in a strong magnetic field \cite{Sikivie_experimental_1983}. Although very sensitive, these experiments are highly model-dependent, relying on unverified assumptions about the local density and distribution of dark matter. Based on several N-body simulations, the local dark matter distribution is expected to diverge from the currently widely assumed standard halo model \cite{eggemeier_minivoids_2022, vogelsberger_streams_2011}. This prediction severely limits the effective sensitivity of haloscope measurements. Experiments using the light-shining-through-the-wall (LSW) technique are model-independent and intrinsically broadband, thus covering a large portion of the available axion parameter space \cite{mueller_detailed_2009}. 
However, they require conversion of photons into axions and vice versa, which in turn limits their sensitivity.

Interferometric experiments \cite{tam_production_2012} have emerged as an innovative approach in the search for axions due to their exceptional sensitivity to very small changes in amplitude, phase, and polarization. The recently proposed Weak Interacting Slim Particle searches on a Fiber Interferometer (WISPFI) experiment \cite{wispfi_2024} benefits from resonant conversion in a fiber-based interferometer. WISPFI allows for adjusting the refractive index by varying the gas pressure in the hollow core of a photonic crystal fiber, so that the resonant condition can be tuned to reach the Dine-Fischler-Srednicki-Zhitnitsky (DFSZ-)predicted photon-axion coupling. 

Motivated by this approach, a novel broadband experiment called \textbf{W}eak Interacting Slim Particle \textbf{INTER}ferometer (WINTER) is proposed in this paper. WINTER uses a free-space Mach-Zehnder interferometer (MZI) to detect the photon flux reduction due to the conversion of photons to axions according to the Primakoff effect \cite{Primakoff_1935}. The key element of WINTER is the placement of the sensing arm of the interferometer in vacuum and under a strong magnetic field, which, together with the implementation of a Fabry-P\'erot cavity (FPC), boosts the expected sensitivity. 

This approach makes WINTER a unique broadband, model-independent, and highly sensitive experiment capable of detecting axion-like particles (ALPs)\footnote{The experiment is also sensitive to photon-scalar coupling.} and axions beyond the theoretical axion band and covering a large unexplored portion of the axion parameter space. This region partially aligns with the predictions from lattice QCD, which estimate that axions making up a portion of dark matter in a post-inflationary scenario would have masses in the range \SIrange{50}{1500}{\micro\eV} \cite{Borsanyi_2016_mass}.

\begin{figure*}[!htb]
    \centering
    \includegraphics[width=\linewidth]{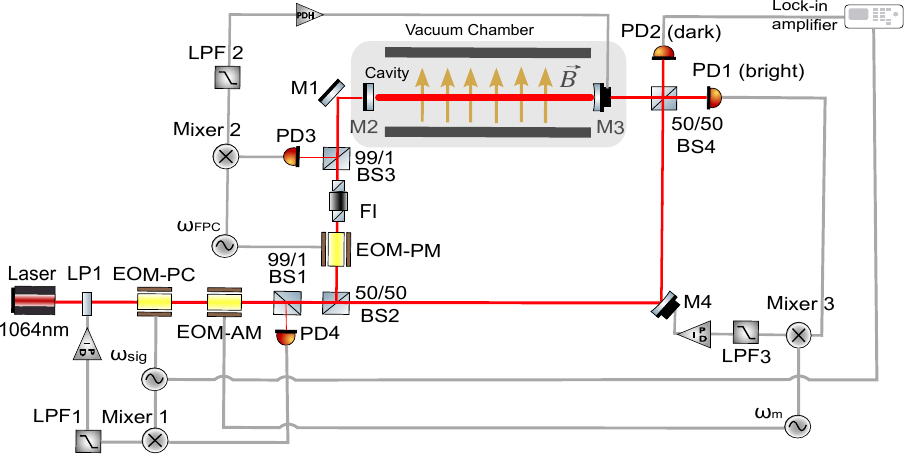}
    \caption{Schematic view of the experimental setup of WINTER using a free-space MZI for broadband detection. In red, the laser beam in free-space is shown. The sensing arm of the interferometer is placed inside a vacuum chamber with a FPC of \SI{10}{\meter} length integrated in a dipole magnet with  \SI{9}{\tesla} of the same length. The acronyms correspond to: electro-optical modulator (EOM) with amplitude (EOM-AM), phase (EOM-PM), and polarization (EOM-PC) modulation,  Faraday isolator (FI), linear polarizer (LP), beam-splitter (BS),  photo-detector (PD), mirror (M), and low-pass filter (LPF). More details about the setup are given in Sect.~\ref{sec:setup_overview}.}
    \label{fig:general_setup}
\end{figure*}

\section{Photon-axion mixing in vacuum}
\label{sec:photon_alps_mixing}
In the following, the axion is treated as a classical pseudo-scalar field, $a$, that mixes with the electric field in the presence of an external magnetic field. The mass of the axion, $m_\mathrm{a}$, is assumed to be much smaller than the energy of the photons as expressed by their angular frequency, $\omega_{\gamma}$, so that the condition $|m_\mathrm{a}|^2 \ll \omega_{\gamma}^2$ is always satisfied. In the plane-wave approximation, a Helmholtz-like equation for the state $\psi(z)=\begin{pmatrix} E(z) \\ a(z) \end{pmatrix}$
can be derived \cite{raffelt_mixing_1988}:

\begin{equation}
\label{eq:maxwell-axion-coupled}
   \left[\partial^{2}_{z}+M^2\right]\psi(z)=0,
\end{equation} 
where $z$ is the direction of propagation. 
$M^2$ is the mixing matrix:
\begin{equation}
\label{eq:mixing_matrix}
M^2=\frac{1}{2\omega_\gamma}\begin{pmatrix} k^2_{\gamma} & -G \\ -G & k^2_{a} \end{pmatrix},
\end{equation} 
where $k_{\gamma}^{2}=\omega_{\gamma}^{2}-m_{\gamma}^{2}$ and $k_{a}^{2}=\omega_{a}^{2}-m_{a}^{2}$ are the photon and axion momenta respectively 
and $G=g_{a\gamma\gamma}B_\mathrm{ext}/2$ is the mixing energy which is proportional to the external magnetic field and the coupling between axion and photon $g_{a\gamma\gamma}$.

The resulting probability of photon-to-axion conversion, $P_{\gamma \rightarrow a}$ is calculated in a  base rotated by an angle $\theta$ such that:
\begin{equation}
\label{eq:pconv}
    P_{\gamma \rightarrow a}= \sin^{2}(2\theta) \sin^{2}\left(k_\mathrm{osc} z\right),
\end{equation} 
where $k_\mathrm{osc}=\sqrt{G^2 + \left(\frac{m_a^2}{4\omega_\gamma}\right)^2}$ is the oscillation wavenumber under the approximation $\omega_\gamma \approx \omega_a \gg m_a$ and $m_\gamma =0$ (vacuum conditions). The mixing angle, $\theta$, calculated by diagonalizing the mixing matrix, is given by
\begin{equation}
\label{eq:mixing_angle}
  \tan(2\theta)=\frac{4\omega_\gamma G}{k_\gamma^2-k_a^2},
\end{equation} 
and the conversion probability can be re-written as
\begin{equation}
\label{eq:pconv_vac}
    P_{\gamma \rightarrow a} = \frac{G^{2}}{G^{2} + \frac{m_{a}^4}{16\omega_{\gamma}^2}} \sin^{2}\left(\sqrt{G^{2} + \frac{m_{a}^4}{16\omega_{\gamma}^2}} z\right).
\end{equation}

For $k_\mathrm{osc}z\ll 1$, the sine function can be expanded, with the resulting (mass-independent) conversion probability  
\begin{equation}
\label{eq:pconv_vac_linear}
    P_{\gamma \rightarrow a} \simeq (Gz)^2=\frac{g_{a\gamma\gamma}^2 B_\mathrm{ext}^2}{4} z^2.
\end{equation} 



\section{Interferometric detection}
\label{sec:interferometric_detection}

The principle behind detecting photon-axion conversion in a MZI has already been discussed in detail for the WISPFI experiment \cite{wispfi_2024}.
The experimental scheme of WINTER is shown in Fig.~\ref{fig:general_setup}. WINTER uses an electro-optical modulator (EOM) for modulating the laser beam in amplitude (EOM-AM) while switching between the two orthogonal states of linear polarization passing through a Pockels cell (EOM-PC). The amplitude modulation is performed using a sinusoidal-amplitude modulated electric field:
\begin{equation}
\label{eq:mod_field}
    \mathbf{E(t)} = \frac{E_0}{\sqrt{2}}\left( 
    \begin{array}{c} 
    1 + \beta_\mathrm{m} \sin(\omega_m t+\chi)\\ 
    1 + \beta_\mathrm{m} \sin(\omega_m t)
    \end{array} \right) e^{i \omega_{\gamma} t},
\end{equation}
with an amplitude $E_{0}$, a modulation depth $\beta_\mathrm{m}$, an amplitude-modulation frequency $\omega_{m}$, and a phase offset $\chi\approx \pi/2$ between the two modulation signals.
The external magnetic field, $B_\mathrm{ext}$, is aligned to be perpendicular to the direction of the wave propagation and parallel to one of the linear polarization modes of the laser beam. The sensing and reference arm lengths of the interferometer ($L_\mathrm{sen}$ and $L_\mathrm{ref}$) are chosen to differ by  $\Delta L=L_\mathrm{sen}-L_\mathrm{ref}$. The total length is given by $L=(L_\mathrm{sen}+L_\mathrm{ref})/2$. 

The optical path length difference $\Delta L$ required for destructive interference at the dark fringe corresponds to a large multiple of half-wavelengths.  Experimentally speaking, imperfections in alignment, noise, and locking stability introduce a small deviation from the ideal condition. Therefore, the actual path length difference can be written as
\begin{equation}
\Delta L = (2n+1)\frac{\lambda_{\gamma}}{2} + \sigma_{\gamma},
\label{eq:path_length_difference_main}
\end{equation} where $n \in \mathbb{Z}$, $\lambda_{\gamma}$ is the laser wavelength and $\sigma_{\gamma}$ represents the uncertainty in locking on the dark fringe. 
Applying $\Delta L$ for the destructive interference condition, we obtain the residual phase deviation for the carrier:
\begin{equation}
    k_{\gamma} \Delta L = (2n+1)\pi + \alpha_{\gamma},
\label{eq:carrier_phase_error_main}
\end{equation} 
with $k_{\gamma}={2\pi}/{\lambda_{\gamma}}$ expressing the laser wavenumber, and $\alpha_{\gamma} = k_{\gamma} \sigma_{\gamma}$ quantifying the phase error from imperfect locking. The respective destructive interference condition for amplitude-modulated sidebands is then given by
\begin{equation}
k_m \Delta L = (2p+1)\pi + \alpha_m,
\label{eq:sideband_phase_error_main}
\end{equation}
where $\alpha_m=\frac{k_m\alpha_{\gamma}}{k_{\gamma}}$ is the sideband phase error, $k_m = \omega_m / c$ is the modulation wavenumber, $c$ is the speed of light, and $p \in \mathbb{Z}$ (with $p \ll n$ decoupling effectively the two conditions in Eqs.  \ref{eq:carrier_phase_error_main} and \ref{eq:sideband_phase_error_main}). Since $k_m \ll k_{\gamma}$, the sideband phase error, $\alpha_m$, can be neglected. In the presence of an external magnetic field and non-vanishing photon-axion conversion probability $P_{\gamma\rightarrow a}$, a modulated signal is expected at the dark port originating from the combined interference of the sidebands with each other and with the carrier (for the full derivation see Appendix \ref{sec:interferometric_principle}).

To lock on, for example, the upper sideband at $\omega_{\gamma} + \omega_m$, we first demodulate the resulting power from port 2 (dark port) at the amplitude modulation frequency $\omega_m$. This demodulation scheme is performed using lock-in amplification by mixing the dark port signal with a local oscillator (LO) at $\omega_m$ following a homodyne procedure (see Appendix \ref{sec:demodulation_process} for details), which stabilizes the interferometer at a fixed working point in amplitude. Terms arising from the self-interference of the carrier and sidebands are omitted, as their contributions are negligible after demodulation and averaging the power over long time.  
Once the interferometer is locked, the signal itself is demodulated at the polarization modulation frequency, $\omega_{\rm sig}$, to track the slow amplitude variations induced by the polarization changes. In both demodulation steps, the output of the mixer is passed through a low-pass filter with a bandwidth much lower than the respective modulation frequency, ensuring that only slow or dc variations of the signal are retained. The time-averaged output power of the signal after the mixer is then

\begin{equation}
\label{eq:power_mixing_avarage_main}
    P_{\text{av}} = \frac{1}{2} P_{\text{tot}} \, \beta_\mathrm{m} \, P_{\gamma \rightarrow a},
\end{equation} where $P_{\text{tot}}$ is the input power of the laser. The effect of the polarization modulation through the EOM-PC is accounted by time-averaging the conversion probability (for more details see section \ref{sec:demodulation_process}).

\section{Experimental setup}
\label{sec:setup_overview}

The general setup for the WINTER experiment shown in Fig.~\ref{fig:general_setup} is based on a free-space MZI, as introduced above. One of the arms (sensing arm) is embedded inside the bore of a strong dipole magnet and uses a FPC under high-vacuum conditions, while the other one acts as a reference arm. In the following, we assume a magnet similar to the LHC test dipole magnets, like the one used by the CERN Axion Solar Telescope (CAST) experiment \cite{CAST_new_2017}, with a length of $\approx \SI{10}{\metre}$ and a magnetic field strength of $\approx\SI{9}{\tesla}$.

The FPC increases the total path length of the photons within the magnetic field. Therefore, it increases the conversion probability $P_{\gamma\rightarrow a}$ (Eq.~\ref{eq:pconv_vac_linear}) in a quadratic way and improves the sensitivity because of the effectively larger circulating power. The FPC employed in WINTER comprises of two mirrors (a planar and a concave one) with an aperture radius of \SI{17.5}{\cm} and a high-power continuous wave laser operating at a wavelength of \SI{1064}{\nm}. With this optical scheme, diffraction and clipping losses are minimized to not degrade the finesse of the cavity (see Appendix \ref{appendix:clipping} for details). Given the oscillation wavenumber, $k_{\rm osc}$, defined in Sect.~\ref{sec:photon_alps_mixing} assuming vacuum conditions with the photon-axion mixing energy satisfying $G \ll m_a^2/(4 \omega_\gamma)$, the oscillation length of the axion field is estimated to be:
\begin{equation}
L_{\rm osc} = \frac{\pi}{k_{\rm osc}} \simeq \frac{4 \pi \, \omega_{\gamma}}{m_a^2}  \simeq \SI{10}{\meter} \left(\frac{m_a}{\SI{0.38}{\meV}}\right)^{-2}.
\end{equation}
For the given length of the magnet, the maximum length of the cavity $L_{\rm FPC} \approx \SI{10}{\meter}$, results in a photon-axion conversion taking place up to a mass of $\approx \SI{0.5}{\meV}$.

In the following, we describe the necessary control loops and associated modulations for the experiment:
\begin{itemize}
    \item[A.] Matching the power between the two polarization states (Sect.~\ref{subsec:modulation}), using the polarization modulation at frequency $\omega_{\mathrm{sig}}$ (EOM-PC).
    \item[B.] Locking the FPC at its working point (Sect.~\ref{subsec:cavity_lock}), using phase modulation at frequency $\omega_{\mathrm{FPC}}$ (EOM-PM).
    \item[C.] Locking the interferometer near a dark fringe (Sect.~\ref{subsec:laser_lock}), using amplitude modulation at frequency $\omega_m$ (EOM-AM).
\end{itemize}

\subsection{Power-matching between the polarization states}
\label{subsec:modulation}

A key aspect of the experiment is the modulation of the expected signal from
photon-axion mixing at the frequency $\omega_{\text{sig}}$ (see Eqn.~\ref{eq:mod_field}),
such that it can be reliably distinguished from other instrumental effects. For photons with a lifetime of $\tau_{ph}$ inside the FPC, the condition $\omega_\text{sig}< 2\pi/\tau_{ph}$ has to be satisfied
(see Sect.~\ref{subsec:cavity_lock} for further details). 
The modulation is achieved by changing the polarization of the light before it enters the interferometer. In the proposed setup (see Fig.~\ref{fig:general_setup}), this is implemented in a two-step process. 
First, the laser's linearly polarized light passes through a Proportional Integral–Derivative(PID)-controlled motorized linear polarizer (LP1), which is set to an angle close to 45 deg relative to the initial polarization direction, such that the resulting field is shared equally between the two linear polarization states. 
In a second step, the Pockels cell (EOM-PC) is driven at the frequency $\omega_\text{sig}$ to create a modulated signal as given by Eq.~\ref{eq:mod_field}. To maintain an equal share of power in the two polarization modes, a beam splitter (99/1 BS1) is used to read out the error signal with a photodiode (PD4), which feeds into a PID loop that changes the direction of LP1. 

A free-space beam splitter (BS1) is then used to split the modulated beam equally and redirect it to the two interferometer arms. 

\subsection{Cavity locking}
\label{subsec:cavity_lock}
The FPC used in the WINTER experiment consists of a plane mirror (M2) and a concave piezo-driven mirror (M3) with relative reflectivity $R$ separated by the distance $L_\mathrm{FPC}=\SI{10}{\metre}$ as shown in Fig.~\ref{fig:general_setup}. The resulting finesse $\mathcal{F}$ of the FPC is given by:
\begin{equation}
\label{eq:finesse_FPC}
  \mathcal{F} = \frac{\pi \sqrt{R}}{1 - R} = \frac{FSR}{\Delta f},
\end{equation} 
which is the ratio between the free-spectral range ($FSR=c/2L_\mathrm{FPC}=\SI{15}{\MHz}$) and the linewidth $\Delta f$ of the cavity modes. For the planned experiment, a transmission loss value of $1-R=\SI{30}{ppm}$ is assumed for each mirror, which accounting for diffraction and clipping losses, it translates into a finesse of $\mathcal{F}\approx 10^5$ and a linewidth of $\Delta f = \SI{150}{\hertz}$. The effects of diffraction in the conversion region and thermal load in the mirrors are discussed in the Appendixes Sect.~\ref{appendix:clipping}.


The locking of the FPC involves adjusting its length using a piezo-driven mirror (M3) to match 
$L_{\mathrm{FPC}} = N\frac{\lambda_\gamma}{2}$ where $N$ corresponds to the cavity mode number assuming vacuum conditions. In this case, the FPC becomes transparent for the incoming laser, no light is reflected, and the circulating power reaches its maximum value.  

The position of the piezoelectric mirror M3 is adjusted by minimizing the sideband signal reflected from the mirror M2
and measured with the photodiode PD3. The sideband is generated by modulating the phase of the carrier with the EOM-PM, following the well-established Pound-Drever-Hall locking scheme \cite{PDH_technique_1983}. A Faraday isolator (FI) is placed after the BS2 to prevent any back reflections from the FPC. The modulation frequency driving the EOM-PM for locking the cavity is then chosen to be half of the FSR at $f_\text{FPC}=\omega_{\text{FPC}}/{(2\pi)}=\SI{7.5}{\MHz}$ so that it does not interfere with the modulation of the signal and the stabilization of the interferometer.



The modulation frequency, $\omega_{\mathrm{sig}}$, of the EOM-PC has to be chosen to be sufficiently small to allow the power in the FPC to build up during the modulation period. 
This timescale should be larger than the photon lifetime:
\begin{equation}
\label{eq:photon_lifetime}
\tau_{ph} = \frac{1}{FSR(1-R)}.
\end{equation}
Therefore, the resulting frequency $f_{\text{sig}} = \omega_{\text{sig}}/(2\pi)$ should fulfill the requirement
\begin{equation}
    f_{\text{sig}} < FSR(1-R)\approx \SI{480}{\hertz}.
\end{equation}
The final choice of $f_\mathrm{sig}$ will be optimized to strike a balance between the required power build-up in the cavity and the low-frequency noise. The frequency could be swept to verify that the response is flat in the relevant frequency range.
We settle here for a value of $f_\text{sig}\approx \SI{20}{\hertz}$ where the average power in the polarization mode parallel to the magnetic field is $\approx  39\%$ and a modulation index between the two polarization modes of $\beta_\mathrm{sig}\approx 1$.




Note that the input laser beam requires a linewidth smaller than the FPC linewidth to guarantee an optimal coupling. For this experiment, the former will be on the order of \SI{0.1}{kHz} \cite{toptica_private}. This ensures that the coherence length of the laser is sufficiently large to match the total optical path length of the FPC.

\subsection{Interferometer stabilization}
\label{subsec:laser_lock}

The interferometer needs to be phase-locked at a working point, which, for WINTER, is chosen to be close to $0.01\%$ of the dark fringe power for PD2. This reduces the corresponding shot noise from the laser and improves the sensitivity (see also Sect.~\ref{sec:sensitivity}).\\
The lock is achieved by driving the piezo-electric mirror M4 to compensate phase changes between the two arms, including residual phase contributions from the optical elements, which are tracked using the error signal generated from PD1, demodulated at the frequency $\omega_m$, the frequency at which the EOM-AM modulates the amplitude of the carrier. The amplitude-modulation frequency is chosen sufficiently high, $f_m = \omega_m/(2\pi) \approx \SI{1}{\kilo\hertz}$, to avoid overlap with the polarization modulation frequency $f_\text{sig} \approx \SI{20}{\hertz}$ and to reduce the impact of $1/f$ noise from the laser.\\
The axion-induced signal is read out at PD2 once the interferometer is locked near the dark fringe. This is done by demodulating the PD2 signal at the polarization-modulation frequency $\omega_{\rm sig}$, applied via the EOM-PC, which tags the axion-sensitive component of the optical field.

\section{Sensitivity analysis}
\label{sec:sensitivity}

The expected sensitivity of the WINTER setup will be estimated assuming that the MZI is operated close to a dark fringe. The main noise sources are the dark current of the PD2, and the shot noise contribution from the sidebands. The resulting signal-to-noise ratio (SNR) is then given by the ratio of the 
signal power received at the dark port (port 2) from photon-axion conversion (see Eqn.~\ref{eq:power_mixing_avarage_main}) and the noise power contributed by the PD2 and by the shot-noise of the laser in the dark fringe. 

For an InGaAs photodiode, the noise-equivalent power from the dark current  ($\mathrm{NEP}_\mathrm{PD2}$) amounts to approximately \SI{0.7}{\femto\watt\per\sqrt{\Hz}} \cite{Femto_FWPR20SI}. The estimated contribution of the shot noise from the sidebands is derived from the shot noise from the residual laser signal in the dark fringe integrated over the bandwidth of the sideband. The resulting $\mathrm{NEP}_\mathrm{SN}$ from the shot-noise is estimated:
\begin{equation}
  \text{NEP}_{\text{SN}} = \sqrt{\frac{2\cdot P_{\text{dark} }\cdot E_{\text{ph}}}{\epsilon}} = \SI{73.4}{\pico\watt\per\sqrt{\Hz}},
\end{equation}
where $P_{\text{dark}}\approx 10^{-4}~P_{tot}$ is the residual power received at the photodetector in the dark fringe. $E_{\text{ph}}$ is the photon's energy and $\epsilon$ is the detection efficiency of the photodetector (PD2), which is assumed to be $90\%$.
For a commercial laser with $P_\mathrm{tot}=\SI{130}{\watt}$ and $\lambda_\gamma = \SI{1064}{\nm}$, the effective sideband noise power for a  bandwidth of $\Delta \nu=\SI{10}{\hertz}$ is then:
\begin{equation}
     \mathrm{P}_\text{SN+PD2} = \sqrt{(\text{NEP}_{\text{PD2}}^2 + \text{NEP}_{\text{SN}}^2) \Delta{\nu}} = \SI{232.2}{\pico\watt}  .
\end{equation}

\begin{figure*}[!htb]
    \centering
    \includegraphics[width=0.8\linewidth]{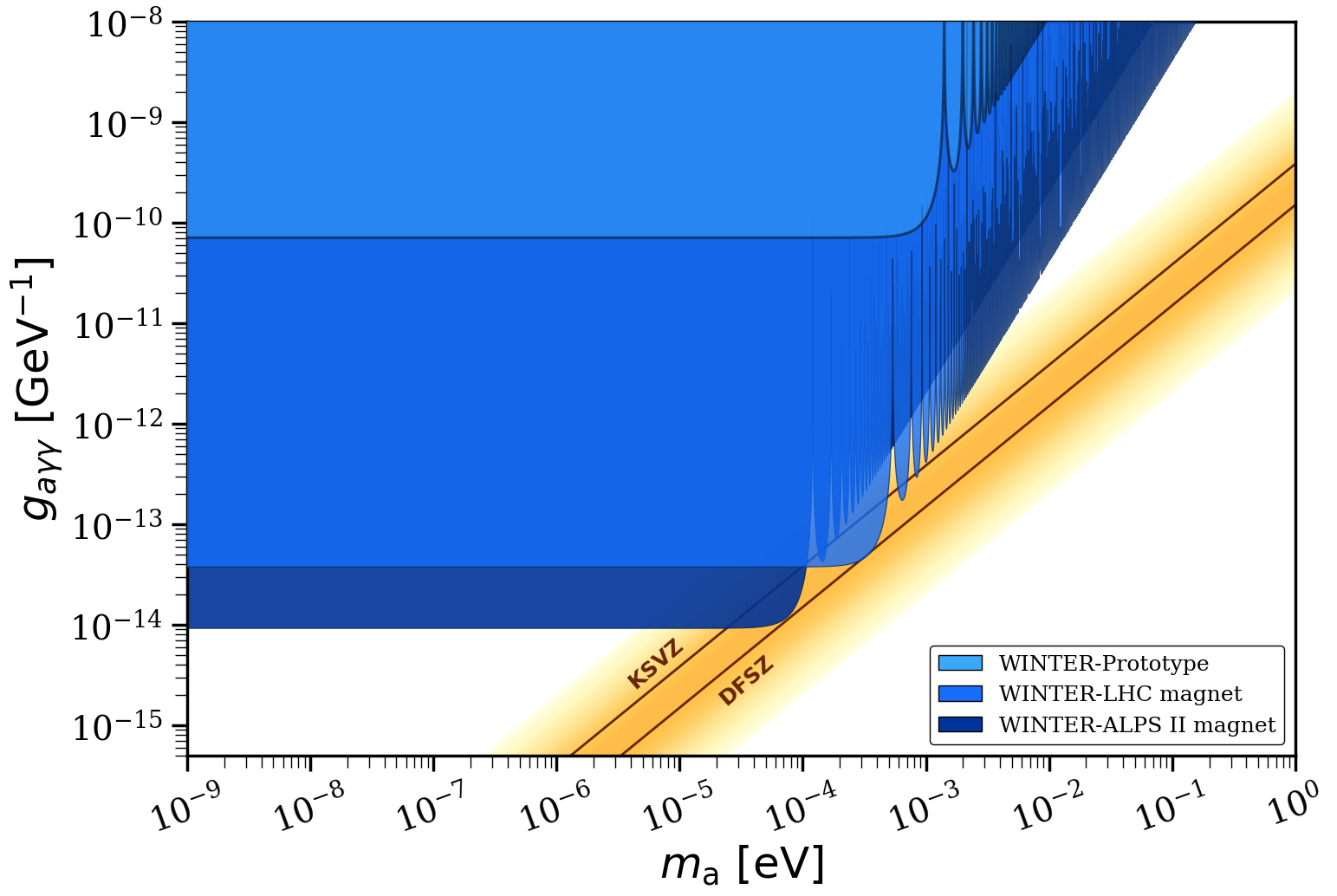}
    \caption{Projected sensitivity for WINTER experiment using an LHC and ALPS-II magnet configurations, as well as a prototype currently under construction at the University of Hamburg. The specific parameters used for the calculation of all three setups are shown in Tab.~\ref{tab:we_comparison}. Note that the differences in the cutoff axion masses are due to the different wavelengths and magnet length along which the FPC is integrated.}
    \label{fig:exclusion_plot} 
\end{figure*}

Considering that the beam traverses a path length of $L_{\text{FPC}}=\SI{10}{\meter}$, in a FPC with $\mathcal{F}=10^5$ finesse, embedded into a \SI{9}{\tesla} LHC-type dipole magnetic field of equal length, and assuming an operation time of $t = \SI{1}{\year}$, the resulting sensitivity for $g_{a\gamma\gamma}$ in the broadband range is given by:
\begin{widetext}
\begin{equation}
\label{eq:sensitivity_baseline_setup}
\begin{aligned}
    g_{a\gamma\gamma} &\approx 3.7\times10^{-14} ~\si{\GeV\tothe{-1}} \left(\frac{\mathrm{SNR}}{3}\right)^{1/2}
    \left(\frac{B_\text{ext}}{\SI{9}{\tesla}}\right)^{-1} \left(\frac{L_{\text{FPC}}}{\SI{10}{\meter}}\right)^{-1}
    \left(\frac{C_\mathrm{dut}}{0.024}\right)^{-1/2}
    \left(\frac{P_\mathrm{tot}}{\SI{130}{\watt}}\right)^{-1/2}
    \left(\frac{\beta_\mathrm{m}}{1}\right)^{-1/2}
    \left(\frac{\beta_\mathrm{\text{sig}}}{1}\right)^{-1}\\
    &\times \left(\frac{t}{\SI{1}{\year}}\right)^{-1/4}
    \left(\frac{\mathrm{NEP_{SN+PD2}}}{\SI{73.4}{\pico\watt\per\sqrt{\Hz}}}\right)^{1/4}
    \left(\frac{\mathrm{\Delta{\nu}}}{\SI{10}{\Hz}}\right)^{1/8}
    \left(\frac{\mathrm{E_{ph}}}{\SI{1.16}{\eV}}\right)^{1/4}
    \left(\frac{\mathcal{F}}{1\times 10^{5}}\right)^{-1/2},
\end{aligned}
\end{equation}
\end{widetext}
where $C_\mathrm{dut}$ is a constant value related to the duty cycle accounting for the 39\% average power of the polarization mode parallel to the magnetic field inside the FPC, as well as the factor 1/16 from Eq.~\ref{eq:power_mixing_average_final} resulting from the time-averaging of the sideband modulation $\omega_m$, the time-averaging of the signal modulation $\omega_\text{sig}$, and the sine expansion of conversion probability. The expected sensitivity for WINTER is shown in Fig.~\ref{fig:exclusion_plot} where the resulting limit on the axion-photon coupling is on the level of $g_{a\gamma\gamma}\gtrsim 3.7\times10^{-14}~\si{\GeV^{-1}}$ for axion masses up to \SI{380}{\micro \eV}. This level of sensitivity allows WINTER to probe the DFSZ model region, representing a significant advancement in exploring the QCD axion parameter space at these masses.

Using instead the string of dipole magnets employed in the ALPS-II experiment, providing a magnetic field strength of
$B_\text{ext}=\SI{5.3}{\tesla}$ over a total magnetic length of $L=\SI{200}{\meter}$~\cite{Bahre2013} and assuming an optical resonator with a reduced finesse of $\mathcal{F}=1.2\times10^{4}$, the sensitivity can be rescaled according to the explicit $B_\text{ext}^{-1}L^{-1}\mathcal{F}^{-1/2}$ dependence in
Eq.~\eqref{eq:sensitivity_baseline_setup}. The finesse is chosen to maximize intracavity power buildup, while accounting for the increased cavity length, which leads to
a longer photon storage time and correspondingly tighter constraints on the cavity linewidth and optical losses. Despite the reduced magnetic field strength compared to the
LHC setup, the substantially increased interaction length leads to a significant improvement in the achievable sensitivity.
Assuming identical experimental parameters, the projected sensitivity on the axion-photon
coupling reaches $g_{a\gamma\gamma} \simeq 9.2\times10^{-15}~\si{\GeV^{-1}},$
with conversion maintained up to an axion mass cutoff of
$m_a \simeq \SI{85}{\micro\eV}$, determined by the total magnetic length.

Presently, a prototype table-top version is already under construction, with a \SI{2}{\watt} \SI{1550}{\nm} laser and a static dipole magnetic field of \SI{1}{\m} in length and about \SI{1.2}{\tesla} in strength. The custom-made magnet panel is constructed from Neodymium permanent magnets separated by \SI{4}{\mm} defined by the required beam width, which is limited by diffraction losses (see Appendix \ref{appendix:clipping}). The Neodymium magnets are aligned to provide a close to uniform dipole magnetic field distribution along the entire length. This prototype setup will use a FPC of \SI{1}{\meter} in length and a finesse of $\mathcal{F}=3\times10^4$. The corresponding FSR and linewidth are \SI{150}{\MHz} and \SI{5}{kHz}, respectively. For a total measuring time of \SI{30}{\day} and a detector with a NEP of \SI{2}{\femto\watt\per\sqrt{\Hz}} \cite{Thorlabs_DET08CL} the expected exclusion limit of $g_{a\gamma\gamma}\gtrsim 7.1\times10^{-11}~\si{\GeV^{-1}}$ for ALP masses up to \SI{996}{\micro \eV} would already improve on existing limits for the axion photon coupling.

\begin{table}[!htb]
\centering
\begin{threeparttable}
\renewcommand{\arraystretch}{1.25}
\resizebox{\columnwidth}{!}{%
\begin{tabular}{|c|c|c|c|}
\hline
\multirow{2}{*}{\textbf{Parameter}} &
\multicolumn{3}{c|}{\textbf{WINTER}} \\
\cline{2-4}
 & \textbf{Prototype} & \textbf{LHC} & \textbf{ALPS-II} \\
\hline
$B_\text{ext}$
& \SI{1.2}{\tesla} 
& \SI{9}{\tesla} 
& \SI{5.3}{\tesla} \\
\hline
$L_{\text{FPC}}$
& \SI{1}{\m} 
& \SI{10}{\m} 
& \SI{200}{\m} \\
\hline
$\lambda_\gamma$ 
& \SI{1550}{\nm} 
& \SI{1064}{\nm} 
& \SI{1064}{\nm} \\
\hline
$\mathcal{F}$ 
& $3 \times 10^4$ 
& $10^5$ 
& $1.2 \times 10^4$ \\
\hline
$t$ 
& 30 days 
& \SI{1}{\year} 
& \SI{1}{\year}  \\
\hline
$P_{\text{tot}}$ 
& \SI{2}{\watt} 
& \SI{130}{\watt} 
& \SI{130}{\watt} \\
\hline
$P_\text{dark}/P_{\text{tot}}$
& $1\%$ 
& $0.01\%$ 
& $0.01\%$ \\
\hline
$\text{NEP}_\text{PD2}$
& \SI{2}{\femto\watt\per\sqrt{\Hz}} 
& \SI{0.7}{\femto\watt\per\sqrt{\Hz}} 
& \SI{0.7}{\femto\watt\per\sqrt{\Hz}} \\
\hline
$\text{NEP}_\text{SN+PD2}$ 
& \SI{75.6}{\pico\watt\per\sqrt{\Hz}} 
& \SI{73.4}{\pico\watt\per\sqrt{\Hz}} 
& \SI{73.4}{\pico\watt\per\sqrt{\Hz}} \\
\hline
$g_{a\gamma\gamma}$\tnote{*} 
& $7.1\times10^{-11}\,\si{\GeV^{-1}}$
& $3.7\times10^{-14}\,\si{\GeV^{-1}}$
& $9.2\times10^{-15}\,\si{\GeV^{-1}}$ \\
\hline
$m_a$\tnote{**} 
& \SI{996}{\micro\electronvolt}
& \SI{380}{\micro\electronvolt}
& \SI{85}{\micro\electronvolt} \\
\hline
\end{tabular}
}
\begin{tablenotes}
    \item[*] Minimum projected mass-independent coupling.
    \item[**] Cut-off axion mass.
\end{tablenotes}
\caption{Comparison of the WINTER-prototype and the full WINTER setup with
the LHC and ALPS-II magnet configurations.}
\label{tab:we_comparison}

\end{threeparttable}
\end{table}

\section{Conclusion}

WINTER is a proposal for a novel experimental setup for broadband searches of axions and ALPs via photon-axion conversion in vacuum. It utilizes a free-space MZI with one arm placed in vacuum and subject to a strong transverse magnetic field, where a FPC enhances the effective circulating power and the total photon path length inside the space pervaded by the magnetic field. The interferometric scheme allows for highly sensitive detection of small amplitude modulations arising from photon-to-axion conversion through the Primakoff effect.

A key advantage of the WINTER setup is its ability to operate near the dark fringe, where shot noise is significantly reduced \cite{Gerberding_interferometers_2022}. This optical noise suppression improves the overall sensitivity, making the system capable of detecting small signals in the presence of dominant background noise. Additionally, the modulation of the polarization state allows the signal to be searched at a selected frequency $\omega_\text{sig}$, distinguishing it from environmental and instrumental effects.

Unlike haloscopes, WINTER does not rely on assumptions about the local dark matter density or its velocity distribution. This approach is model-independent and sensitive to axions regardless of their contribution to the total dark matter abundance, making it a robust probe of photon-axion coupling.

Compared to existing experiments, WINTER offers a unique approach by relying on single-stage photon-to-axion conversion, unlike LSW setups such as Any Light Particle Search II (ALPS II) \cite{alps_demonstration_2020}, which require both conversion and reconversion, thereby reducing the signal strength. Additionally, birefringence-based experiments (e.g., Polarization of Vacuum with LASer (PVLAS) \cite{ejlli_pvlas_2020}) aim to detect axion-induced changes in polarization, however, without using interferometry techniques. Lastly, the CAST experiment \cite{CAST_Micromegas_2024} is used to search for solar axions and therefore relies on axion-generation in the Sun's interior.

WINTER, being broadband and model-independent, is well positioned to explore previously inaccessible regions of axion parameter space, including the theoretically motivated QCD axion band \cite{Borsanyi_2016_mass}. In this regime, WINTER achieves sensitivities down to
$g_{a\gamma\gamma} \simeq 3.7\times10^{-14}~\si{\GeV^{-1}}$ when operated with an LHC-type dipole magnet ($B_\text{ext}=\SI{9}{\tesla}$, $L=\SI{10}{\meter}$), while employing the ALPS-II dipole magnet configuration
($B_\text{ext}=\SI{5.3}{\tesla}$, $L=\SI{200}{\meter}$) improves the sensitivity to
$g_{a\gamma\gamma} \simeq 9.2\times10^{-15}~\si{\GeV^{-1}}$ at the expense of a reduced conversion mass range. We emphasize that our scheme searches for relativistic axions converted at the laser frequency and therefore does not directly probe the non-relativistic halo dark-matter population.


Finally, a prototype table-top version is already under construction at the University of Hamburg. Despite its compact scale, it is expected to demonstrate the feasibility of reaching competitive sensitivity across a broad axion mass range, offering a new and competitive direction in laboratory-based axion searches. \\


\section*{Acknowledgments}

We thank Le Hoang Nguyen for initial discussions and valuable input on the development of this idea. J.M.B. and M.M. acknowledge funding and D.H. support by the Deutsche Forschungsgemeinschaft (DFG, German Research Foundation) under Germany’s Excellence Strategy – EXC 2121 ``Quantum Universe" – 390833306, and through the DFG funds for major instrumentation grant DFG INST 152/824-1. We thank Oliver Gerberding and Martin Tluczykont for their useful comments, discussions, and contributions to the WINTER prototype development. This article is based upon work from COST Action COSMIC WISPers CA21106, supported by COST (European Cooperation in Science and Technology). 

\section*{Data Availability}
The data are not publicly available. The data are available from the authors upon reasonable request.

\begin{widetext}
\section{Appendixes}

\subsection{Interferometric principle}
\label{sec:interferometric_principle}

\subsubsection{Input fields}

The input field at port 1 of WINTER's MZI is represented as a superposition of two linear polarization components differentiated by $\chi=\pi/2$, and modulated at a frequency $\omega_m$. Mathematically, this modulated electric field can be expressed as a Jones vector:
\begin{equation}
\label{eq:mod_field_app}
    \mathbf{E}_{\text{in1}}(t) = E_0\left( 
    \begin{array}{c} 
    1 + \beta_\mathrm{m} \sin(\omega_m t+\chi)\\ 
    1 + \beta_\mathrm{m} \sin(\omega_m t)
    \end{array} \right) e^{i \omega_{\gamma} t},
\end{equation}
while port 2 has no input field:
\begin{equation}
\label{eq:input_field_2}
\mathbf{E}_{\text{in2}}(t) = 
\begin{pmatrix}
0 \\
0
\end{pmatrix}.
\end{equation}

The input 1 field expression from Eq.~\ref{eq:mod_field_app} can be re-expressed in terms of the carrier ($\mathbf{E}_{\text{car}}(t)$) and upper and lower sidebands ($\mathbf{E}_{\text{sid}}(t)$) in Eq.~\ref{eq:mod_field_exp_app} as follows:

\begin{equation}
      \mathbf{E}_{\text{in1}}(t) = \mathbf{E}_{\text{car}}(t) + \mathbf{E}_{\text{sid}}(t)
    \label{eq:input_field_1_car_sid}
\end{equation}
where
\begin{equation}
      \mathbf{E}_{\text{car}}(t) = E_0 e^{i\omega_{\gamma}t} \left(
      \begin{array}{c} 
      1\\
      1
      \end{array} 
    \right),
    \label{eq:mod_field_exp}
\end{equation}

\begin{equation}
      \mathbf{E}_{\text{sid}}(t) =  \frac{E_0\beta_\mathrm{m}}{2i}\left(
      \begin{array}{c} 
      e^{i\left[(\omega_{\gamma}+\omega_m)t+\chi\right]} -
      e^{i\left[(\omega_{\gamma}-\omega_m)t+\chi\right]}\\
      e^{i(\omega_{\gamma}+\omega_m)t} 
      - e^{i(\omega_{\gamma}-\omega_m)t}
      \end{array} 
    \right),
    \label{eq:mod_field_exp_app}
\end{equation}

Following the interferometric scheme from Fig.~\ref{fig:general_setup}, the input fields then go through a 50/50 BS, which is represented by the unitary matrix:
\begin{equation}
\mathbf{U}_{\text{BS}} = \frac{1}{\sqrt{2}}
\mqty(1 & i \\ i & 1)
\end{equation}

Applying that to the input fields $\vb{E}_{\text{in1}}(t)$ (port 1) and $\vb{E}_{\text{in2}}(t)$ (port 2) we obtain the sensing field $\vb{E}_S(t)$ and reference field $\vb{E}_R(t)$:
\begin{equation}
\label{eq:ref_sens_amplitudes}
\begin{pmatrix}
\vb{E}_S \\
\vb{E}_R
\end{pmatrix}
=
\mathbf{U}_{\text{BS1}}
\begin{pmatrix}
\vb{E}_{\text{in1}}(t) \\
\vb{E}_{\text{in2}}(t)
\end{pmatrix}.
\end{equation}

Note that only the polarization component of the beam, which is parallel to the external magnetic field present in the sensing arm, couples to the axion. Therefore, in the following calculations, we will only focus on this case.

\subsubsection{Carrier Amplitude Contribution}

Considering the optical fields at the sensing and reference arms after the first 50/50 BS from Eq.~\ref{eq:ref_sens_amplitudes}, the optical carrier field contributions are estimated, accounting for the photon-axion mixing effect ($P_{\gamma \rightarrow a}$) and the length contributions from each arm, which are given by: 

\begin{equation}
E_{\mathrm{car},S} = \frac{E_{0}}{\sqrt{2}}(1 - P_{\gamma \rightarrow a}) ~e^{i\left[\omega_{\gamma} t - k_{\gamma}\left(L + \frac{\Delta L}{2}\right) \right]},
\end{equation}

\begin{equation}
E_{\mathrm{car},R} = \frac{iE_{0}}{\sqrt{2}} e^{i\left[\omega_{\gamma} t - k_{\gamma}\left(L - \frac{\Delta L}{2}\right) \right]},
\end{equation}
where $L_S$ and $L_R$ represent the optical path lengths of the sensing and reference arms respectively, $L$ is the macroscopical optical path length, $L=\frac{L_R+L_S}{2}$, and $\Delta L=L_R-L_S$ is the path length difference.  

The carrier field components from each arm after recombination from the second 50/50 BS are:
\begin{equation}
E_{\text{car,out1}} = \frac{E_0}{2} e^{i\omega_{\gamma} t} 
    \left\{ e^{-i\left[k_{\gamma} \left(L+\frac{\Delta L}{2}\right)\right]}\left(1-P_{\gamma\rightarrow a}\right)
     - e^{-ik_{\gamma} \left(L-\frac{\Delta L}{2}\right)} \right\},
\label{eq:output_carrier_1}
\end{equation}

\begin{equation}
E_{\text{car, out2}} = i\frac{E_0}{2} e^{i\omega_{\gamma} t} 
    \left\{ e^{-i\left[k_{\gamma} \left(L+\frac{\Delta L}{2}\right)\right]}\left(1-P_{\gamma\rightarrow a}\right) 
     + e^{-ik_{\gamma} \left(L-\frac{\Delta L}{2}\right)}  \right\}.
\label{eq:output_carrier_2}
\end{equation}

The general condition for destructive interference (dark fringe) for the carrier field in a two-path system  based on a MZI is satisfied when the phase difference between the two arms is an odd multiple of $\pi$:
\begin{equation}
    \label{eq:dark_fringe_condition_carrier}
    k_{\gamma}\Delta L = (2n+1)\pi,
\end{equation}
where $n$ is an integer representing the fringe order. For infrared lasers (e.g. $\lambda_\gamma= \SI{1064}{\nm}$), the path length differences involved are macroscopic with $n$ being a large number ($n\sim10^5$). In practice, achieving a perfect destructive interference can become challenging due to adjustment imperfections, locking errors, or noise. The actual path difference can be written as:
\begin{equation}
\label{eq:path_length_difference}
    \Delta L = (2n+1)\frac{\lambda_{\gamma}}{2}+\sigma_{\gamma},
\end{equation} 
where $\sigma_{\gamma}$ is the experimental uncertainty in locking onto the dark fringe. Substituting into the interference condition from Eq.~\ref{eq:dark_fringe_condition_carrier}, we get:

\begin{equation}
\label{eq:dark_fringe_condition_carrier_alpha}
     k_{\gamma}\Delta L = (2n+1)\pi +\alpha_{\gamma},
\end{equation}
where $\alpha_{\gamma} = k_{\gamma}\sigma_{\gamma}$ represents the residual phase error due to imperfect fringe locking.

By applying condition \ref{eq:dark_fringe_condition_carrier_alpha}, the resulting output carrier fields become:
\begin{equation}
E_{\text{car,out1}} = \frac{E_0}{2} e^{i(\omega_{\gamma} t - k_{\gamma} L)} e^{-i\left[(2n+1)\pi +\alpha_{\gamma}\right]} \left[ (1-P_{\gamma\rightarrow a}) + e^{i\alpha_{\gamma}} \right],
\label{eq:output_carrier_1_bright_general}
\end{equation}

\begin{equation}
E_{\text{car,out2}} = i\frac{E_0}{2} e^{i(\omega_{\gamma} t - k_{\gamma} L)} e^{-i\left[(2n+1)\pi +\alpha_{\gamma}\right]} \left[ (1-P_{\gamma\rightarrow a}) - e^{i\alpha_{\gamma}} \right].
\label{eq:output_carrier_2_dark_general}
\end{equation}
where the output port 1 becomes the bright port and port 2 the dark port.

\subsubsection{Sideband Amplitude Contribution}

The upper(+) and lower(-) sideband contribution for the sensing and reference arms is given by: 
\begin{equation}
\begin{split}
E_{\mathrm{sid},(\pm),S} 
= \pm \frac{\beta_\mathrm{m} E_0}{2i\sqrt{2}} (1 - P_{\gamma \rightarrow a}) 
&  ~e^{i\left[(\omega_{\gamma} \pm \omega_m)t -  \phi^{\pm}_S\right]},
\end{split}
\end{equation}

\begin{equation}
\begin{split}
E_{\mathrm{sid},(\pm),R} 
= \pm\frac{\beta_\mathrm{m} E_0}{2\sqrt{2}} 
& ~ e^{i\left[(\omega_{\gamma} \pm \omega_m)t -  \phi^{\pm}_R\right]},
\end{split}
\end{equation}
where the phase contributions for each arm are:
\begin{equation}
    \begin{aligned}
        \phi^{\pm}_S & = (k_{\gamma} \pm k_m)\left(L + \frac{\Delta L}{2}\right) \\
        \phi^{\pm}_R &= (k_{\gamma} \pm k_m)\left(L - \frac{\Delta L}{2}\right)
    \end{aligned}
    \label{eq:sideband_phase_arms}
\end{equation}
The total sideband contribution is described by the sum of the two sidebands:
\begin{equation}
E_{\mathrm{sid}} = E_{\mathrm{sid}(+)} + E_{\mathrm{sid}(-)}.
\end{equation}

After recombining the beams through the second 50/50 beam splitter, the field output at port 1 and port 2 is given by:
\begin{equation}
E_{\text{sid, out1}} = i\frac{\beta_\mathrm{m} E_0}{4} 
    \left\{e^{i\psi_+} \left[-e^{-i\xi_+}\left(1-P_{\gamma\rightarrow a}\right)+e^{i\xi_+} \right]
     - e^{i\psi_-} \left[-e^{-i\xi_-}\left(1-P_{\gamma\rightarrow a}\right)+e^{i\xi_-} \right]\right\}
\label{eq:output_sideband_1}
\end{equation}

\begin{equation}
E_{\text{sid, out2}} = \frac{\beta_\mathrm{m} E_0}{4} 
    \left\{e^{i\psi_+} \left[e^{-i\xi_+}\left(1-P_{\gamma\rightarrow a}\right)+e^{i\xi_+} \right]
     - e^{i\psi_-} \left[e^{-i\xi_-}\left(1-P_{\gamma\rightarrow a}\right)+e^{i\xi_-} \right]\right\},
\label{eq:output_sideband_2}
\end{equation}
where:
\begin{equation}
\psi_{\pm} = (\omega_{\gamma} \pm \omega_m)t - (k_{\gamma} \pm k_m)L,
\label{eq:sideband_phases_L}
\end{equation}

\begin{equation}
\xi_{\pm} = (k_{\gamma} \pm k_m)\frac{\Delta L}{2}.
\label{eq:sideband_phases_DeltaL}
\end{equation}

For the sidebands modulation, destructive interference is met when $(k_{\gamma} + k_m)\Delta L / 2 =\pi$  for the upper (+) sideband and $ (k_{\gamma} - k_m)\Delta L / 2 = 0$ for the lower (-) sideband. Analogous to the carrier interference condition from Eq.~\ref{eq:dark_fringe_condition_carrier_alpha}, we derive:

\begin{equation}
    \label{eq:modulated_condition_wavenumber_alpha}
    k_m \Delta L = (2p+1)\pi + \alpha_m,
\end{equation} 
where $p \in \mathbb{Z}~(\mathcal{O}(1))$ and $\alpha_m=\frac{k_m\alpha_{\gamma}}{k_{\gamma}}$. Since $k_m \ll k_{\gamma}$, the residual phase error from the sideband locking, $\alpha_m$, is $\sim6 - 8$ orders of magnitude smaller than the one for the carrier $\alpha_{\gamma}$, which can be assumed to be dominant, and $\alpha_m$ can therefore be neglected. The resulting phase contributions from Eqs.~\ref{eq:sideband_phases_L}-\ref{eq:sideband_phases_DeltaL} become:
\begin{equation}
\begin{aligned}
\psi_{\pm} = (\omega_{\gamma} \pm \omega_{m})t - \frac{L}{\Delta L} \left\{ \left[2n + 1 \pm (2p + 1)  \right]\pi + \alpha_{\gamma} \right\}
\end{aligned}
\label{eq:sideband_phases_psi}
\end{equation}

\begin{equation}
\begin{aligned}
\xi_{\pm} = \frac{\left[2n+1\pm (2p+1)\right]\pi + \alpha_{\gamma} }{2}
\end{aligned}
\label{eq:sideband_phases_phi}
\end{equation}

\subsubsection{Power Output Estimation}

To estimate the power we expand the complex expressions using Eqs.~\ref{eq:output_carrier_1_bright_general}-\ref{eq:output_carrier_2_dark_general} and Eqs.~\ref{eq:output_sideband_1}-\ref{eq:output_sideband_2} with the phase values from Eqs.~\ref{eq:sideband_phases_psi}-\ref{eq:sideband_phases_phi}:

\begin{equation}
\begin{aligned}
P_{\text{out1}} &=  E_{\text{car,out1}}^* \cdot E_{\text{car,out1}} +  E_{\text{car,out1}}^* \cdot E_{\text{sid,out1}} + E_{\text{car,out1}} \cdot E_{\text{sid,out1}}^* + E_{\text{sid,out1}}^* \cdot E_{\text{sid,out1}} =\\
&= \frac{1}{4} E_0^2 \biggl\{2 + (-2 + P_{\gamma \rightarrow a}) P_{\gamma \rightarrow a} - 2 (-1 + P_{\gamma \rightarrow a}) \cos\alpha_{\gamma} \\
& \qquad\qquad  + \beta_\mathrm{m}^2 \Bigl[2 + (-2 + P_{\gamma \rightarrow a}) P_{\gamma \rightarrow a} + 2 (-1 + P_{\gamma \rightarrow a}) \cos\alpha_{\gamma}\Bigr] \cos^2\left[ \omega_m t - \frac{(2p+1)\pi}{\Delta L} L\right] \\
&\qquad\qquad - 2 (-1)^p \beta_\mathrm{m} (-2 + P_{\gamma \rightarrow a}) P_{\gamma \rightarrow a} \cos\left[ \omega_m t - \frac{(2p+1)\pi}{\Delta L} L\right]\biggr\},
\end{aligned}
\label{eq:output_power_bright}
\end{equation}

\begin{equation}
\label{eq:power_output_dark_port}
\begin{aligned}
P_{\text{out2}} &= E_{\text{car,out2}} ^* \cdot E_{\text{car,out2}} +  E_{\text{car,out2}}^* \cdot E_{\text{sid,out2}} + E_{\text{car,out2}} \cdot E_{\text{sid,out2}}^* + E_{\text{sid,out2}}^* \cdot E_{\text{sid,out2}} =\\
&= \frac{1}{4} E_0^2 \biggl\{2 + (-2 + P_{\gamma \rightarrow a}) P_{\gamma \rightarrow a} + 2 (-1 + P_{\gamma \rightarrow a}) \cos\alpha_{\gamma} \\
&\qquad \qquad  + \beta_\mathrm{m}^2 \Bigl[2 + (-2 + P_{\gamma \rightarrow a}) P_{\gamma \rightarrow a} - 2 (-1 + P_{\gamma \rightarrow a}) \cos\alpha_{\gamma}\Bigr] \cos^2\left[ \omega_m t - \frac{(2p+1)\pi}{\Delta L} L\right] \\
&\qquad\qquad  - 2 (-1)^p \beta_\mathrm{m}(-2 + P_{\gamma \rightarrow a}) P_{\gamma \rightarrow a} \cos\left[ \omega_m t - \frac{(2p+1)\pi}{\Delta L} L\right]\biggr\}.
\end{aligned}
\end{equation}

For perfect dark fringe conditions ($\alpha_{\gamma}=0$) in both the carrier and the upper (+) modulated sideband, considering $p=0$, the output powers from each port can be simplified to:
\begin{equation}
P_{\text{out1}} = \frac{1}{4} E_0^2 \biggl[(-2 + P_{\gamma \rightarrow a})^2 + \beta_\mathrm{m}^2  P_{\gamma \rightarrow a}^2  \cos^2\left( \omega_m t - k_m L\right) - 2 \beta_\mathrm{m}(-2 + P_{\gamma \rightarrow a}) P_{\gamma \rightarrow a}  \cos\left( \omega_m t - k_m L\right)\biggr]
\label{eq:output_power_bright_alpha_zero}
\end{equation}

\begin{equation}
P_{\text{out2}} = \frac{1}{4} E_0^2 \biggl[P_{\gamma \rightarrow a}^2  + \beta_\mathrm{m}^2  \left(-2 + P_{\gamma \rightarrow a}\right)^2  \cos^2\left( \omega_m t - k_m L\right) - 2 \beta_\mathrm{m} (-2 + P_{\gamma \rightarrow a}) P_{\gamma \rightarrow a} \cos\left( \omega_m t - k_m L\right)\biggr].
\label{eq:output_power_dark_alpha_zero}
\end{equation}
It is noted that, as expected, the cross-product between the carrier and the sidebands does not depend on $\alpha_{\gamma}$ as it cancels out. However, $\alpha_{\gamma}$ still has an impact on the carrier and sideband self-interference power contributions.

\subsection{Demodulation Process}
\label{sec:demodulation_process}

As an example, to lock to the upper sideband at $\omega_{\gamma} + \omega_m$, the resulting power from port 1 (bright port) needs to be demodulated at a frequency $\omega_m$. Note that the output powers resulting from the self-interference of the carrier and sidebands (first and second terms of Eq.~\ref{eq:output_power_bright}), respectively, are omitted since their contribution is negligible after demodulating and averaging their power over long periods. In addition, we can approximate by considering only the first-order terms accounting for $P_{\gamma\rightarrow a}\ll1$:
\begin{equation}
    P_{cross,out1} (t) = E_0^2 \beta_\mathrm{m} P_{\gamma\rightarrow a} \cos(\omega_m t - k_m L)
\label{eq:Power_dark_port_cross_term}.
\end{equation} 
The demodulation approach uses a lock-in amplification, by mixing the signal from the bright port with a LO at the same frequency $\omega_m$, following a homodyne procedure. Regarding that, the output of the mixer is:
\begin{equation}
\label{eq:power_mixing}
    P_{\text{mix}}(t) =  P_{cross,out1} (t) \cos(\omega_m t + \phi_{\text{LO}}),
\end{equation} 
where $\phi_{LO}$ is the phase of the LO. The resulting signal power then passes through a low-pass filter (LPF) to remove high-frequency components ($2\omega_m$). The time-averaged output power of the mixer assuming $\phi_{\text{LO}} = 0$ is:
\begin{equation}
\label{eq:power_mixing_average}
    P_{\text{av}} = \frac{1}{2} P_{\text{tot}} \, \beta_\mathrm{m} \, P_{\gamma \rightarrow a},
\end{equation}
where $P_{\text{tot}} = E_0^2$ is the input power of the laser. 

After the interferometer is locked, the axion-induced signal is recovered by demodulating the dark port 2 output at twice the polarization-modulation frequency $2\omega_{\rm sig}$, driven via the EOM-PC. Note in this case that the photon-to-axion conversion probability, assuming the application of an external magnetic field, is proportional to the square of the interaction term, $(\mathbf{E}\cdot\mathbf{B_\text{ext}})^2$. If the electric field is modulated along the direction of the magnetic field through the EOM-PC, the dot product can be written as
\begin{equation}
\mathbf{E} \cdot \mathbf{B_\text{ext}} = |\mathbf{E}| \, |\mathbf{B_\text{ext}}| \, \beta_{\rm sig} \cos(\omega_{\rm sig} t),
\label{eq:E_B_dot}
\end{equation}
where $\beta_{\rm sig}$ is the modulation depth.  Squaring Eq.~\ref{eq:E_B_dot} leads to a time-dependent $P_{\gamma \rightarrow a}(t)$:
\begin{equation}
P_{\gamma \rightarrow a}(t)=(\mathbf{E}\cdot\mathbf{{B_\text{ext}}})^2 \propto B_{\text{ext}}^2 \, \beta_{\rm sig}^2 \cos^2(\omega_{\rm sig} t).
\end{equation}
Averaging over time gives
\begin{equation}
\label{eq:time_dependent_probability}
P_{\gamma \rightarrow a}\propto B_{\text{ext}}^2 \, \frac{1}{2} \beta_{\rm sig}^2.
\end{equation}
so the time-averaged conversion probability acquires an additional factor $P_{\gamma \to a} \propto \frac{1}{2} \, \beta_{\rm sig}^2$. Therefore, the resulting output power from Eq.~\ref{eq:power_mixing_average}, while using Eqs.~\ref{eq:time_dependent_probability} and \ref{eq:pconv_vac_linear}, becomes
\begin{equation}
\label{eq:power_mixing_average_final}
    P_{\text{av}} = \frac{1}{16} P_\text{tot} \, \beta_\mathrm{m} \, \beta{_{\rm{sig}}^2} \, g_{a\gamma\gamma}^2 \, B_{\text{ext}}^2 \, z^2.
\end{equation}

\subsection{Optical Cavity Losses}
\label{appendix:clipping}

To effectively achieve a high finesse of $\mathcal{F}\approx 10^5$, it is critical to have optimal mirror dimensions and beam aperture to keep the clipping and diffraction losses at a minimum. When accounting for multiple loss mechanisms in the FPC (clipping, diffraction, absorption), the total loss $\delta_{\text{total}}$ is the sum:
\begin{equation}
    \delta_{\text{total}} = \delta_\text{clip} + \delta_\text{diff} + \delta_\text{trans,1} + \delta_\text{trans,2},
\end{equation}
where the round-trip losses caused by the mirror clipping and diffraction are given by $\delta_\text{clip}$ and $\delta_\text{diff}$ respectively, and$\delta_\text{trans,1}$, $\delta_\text{trans,2}$ are the transmission losses of the two mirrors. The expression for the Finesse shown in Eq.~\ref{eq:finesse_FPC} can be simplified when the aforementioned losses are small ($\delta_{\text{total}} \ll 1$) \cite{Kogelnik:66}:
\begin{equation}
    \mathcal{F} = \frac{2\pi}{(\delta_\text{clip} + \delta_\text{diff} + \delta_\text{trans,1} + \delta_\text{trans,2})}.
\end{equation}

\subsubsection{Analytical approximation}

Assuming Gaussian beam propagation, the beam expands from its waist at the plane mirror ($w_\text{plane}$) to a larger radius at the concave mirror ($w_\text{conc}(L_{\mathrm{FPC}})$), calculated from the Rayleigh range as follows:
\begin{equation}
\label{eq:Rayleigh_range}
    z_R = \frac{\pi w_{plane}^2}{\lambda_\gamma}
\end{equation}
\begin{equation} 
\label{eq:beam_waist_concave}
    w_\text{conc}(L_{\mathrm{FPC}}) = w_\text{plane}\sqrt{1 + \left(\frac{L_{\mathrm{FPC}}}{z_R}\right)^2}
\end{equation}

Clipping losses arise from the mode mismatch between the beam waist and the effective radius (aperture) of the concave mirror ($a = \SI{17.5}{\cm}$), expressed as:
\begin{equation}
\delta_{\text{clip},\text{conc}} = \exp\left(-\frac{2a^2}{w_\text{conc}^2}\right)
\end{equation}

Since both mirrors truncate the beam with their respective beam waists, the round-trip combination is:
\begin{equation}
\label{eq:rt_clip_exact}
    \delta_{\text{clip}} = 1 - (1 - \delta_{\text{clip},\text{plane}})(1 - \delta_{\text{clip},\text{conc}}),
\end{equation}
which for small losses ($\delta_{\text{clip}}\ll 1$) reduces to the additive approximation:
\begin{equation}
\delta_{\text{clip}} \approx \delta_{\text{clip},\text{conc}} + \delta_{\text{clip},\text{plane}}.
\end{equation}

\subsubsection{Numerical estimation}

In a stable optical cavity, the fundamental Gaussian mode is well confined by the mirrors, but finite apertures and mirror shapes cause a fraction of the light to diffract outside the ideal mode. This effect is not captured by the clipping loss formulation, since even a beam entirely inside the mirror aperture will gradually lose power due to diffraction at each round trip. The Fox--Li method provides an accurate numerical estimation by finding the optimal eigenmode of the cavity via numerical propagation of the field between mirrors with Fresnel diffraction and application of the mirror aperture and curvature at each step \cite{Fox_Li_1961}.  

The Fox--Li method treats an optical cavity as a round-trip propagation operator $\mathcal{M}$ that maps a transverse field $E(x,y)$ at one mirror surface into the field after one round trip, including free-space Fresnel propagation, phase shifts due to mirror curvature, and amplitude masks representing apertures or mirror edges. By applying $\mathcal{M}$ repeatedly to an arbitrary initial field $E_0(x,y)$, the iteration gradually filters out unstable components, and the field converges to the cavity's dominant eigenmode. Mathematically, this corresponds to:
\begin{equation}
E_{n+1} = \mathcal{M} E_n  \xrightarrow{n\rightarrow\infty} E_{n+1}= \eta E_n,
\end{equation}
where $\eta$ is a complex eigenvalue whose magnitude determines the round-trip diffraction loss by:
\begin{equation}
\delta_{\mathrm{diff}} = 1 - |\eta|^2.
\end{equation}

It is noted that the clipping losses average to zero as they are tiny compared to the aperture of the cavity mirrors. For small $w_\text{plane}$ (typically $< a/3$), the clipping losses are increased due to the high beam divergence, while for $w_\text{plane}>a/3$ the clipping losses increase again due to overfilling of the active area of the mirror. The total losses $\delta_\text{tot}$ for the FPC of the WINTER setup are shown in Fig.~\ref{fig:diffraction_losses} as a function of the beam waist of the plane mirror ($w_\text{plane}$).

\begin{figure}[!htb]
    \centering
    \includegraphics[width=0.7\linewidth]{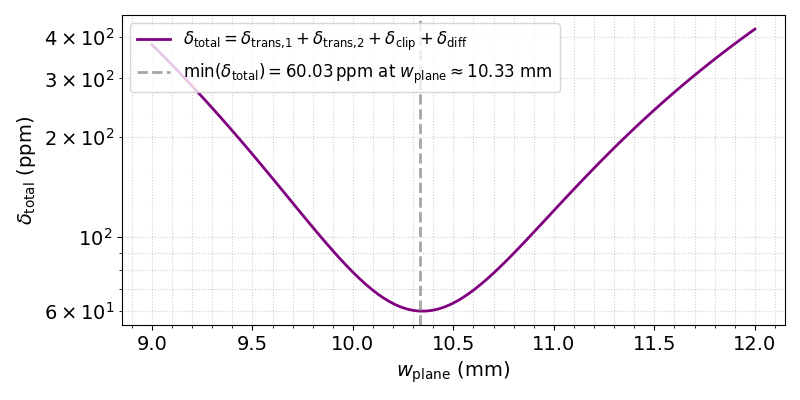}
    \caption{Representation of the total losses in ppm (in blue) for WINTER experiment as a function of the beam waist at the plane mirror (M2). The selected parameters are shown in Tab.~\ref{tab:we_fpc_comparison}.}
    \label{fig:diffraction_losses}
\end{figure}

The impact of the total losses on the finesse of the cavity is shown in Fig.~\ref{fig:finesse_and_wconc_vs_wplano} as a function of the beam waist at the plane mirror. For WINTER, the optimal beam waist at the plane mirror is about $\SI{10.33}{\mm}$ with a concave mirror's beam waist of equal value. It is noted that a FPC with similar properties has already been implemented in vacuum conditions and locked from other experiments (see e.g. ALPS II \cite{alps_demonstration_2020}), demonstrating the feasibility of this approach.

\begin{figure}[!htb]
    \centering
    \includegraphics[width=0.7\linewidth]{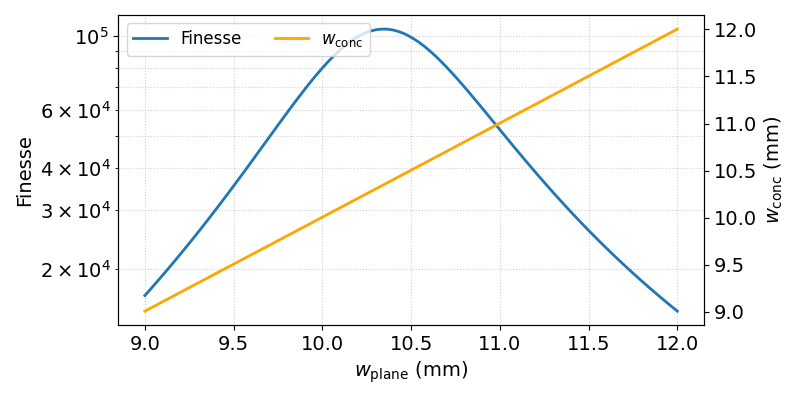}
    \caption{Representation of the finesse variation (in blue) for WINTER experiment as a function of the beam waist at the plane mirror (M2), accounting for diffraction and clipping losses. At the secondary y-axis, the beam waist at the concave mirror (M3) is also shown (in orange). The selected FPC details are shown in Tab.~\ref{tab:we_fpc_comparison}.}
    \label{fig:finesse_and_wconc_vs_wplano}
\end{figure}

Our calculations primarily focus on the configuration using an LHC dipole magnet. 
For comparison, we also estimate the parameters for a setup employing the ALPS-II string of dipole magnets. The relevant FPC parameters and corresponding losses for WINTER, with both LHC and ALPS-II dipole magnets, as well as for the WINTER-prototype apparatus, are summarized in Tab.~\ref{tab:we_fpc_comparison}.

\begin{table}[!htb]
\centering
\renewcommand{\arraystretch}{1.25}
\begin{tabular}{|l|c|c|c|c|}
\hline
\multirow{2}{*}{\textbf{Parameter}} &
\multirow{2}{*}{\textbf{Notation}} &
\multicolumn{3}{c|}{\textbf{WINTER}} \\
\cline{3-5}
 &  & \textbf{Prototype} & \textbf{LHC} & \textbf{ALPS-II} \\
\hline
Laser power 
& $P_\text{tot}$ 
& \SI{2}{\watt} 
& \SI{130}{\watt} 
& \SI{130}{\watt} \\
\hline
Cavity length 
& $L_{\text{FPC}}$ 
& \SI{1}{\m} 
& \SI{10}{\m} 
& \SI{200}{\m} \\
\hline
Laser wavelength 
& $\lambda_\gamma$ 
& \SI{1550}{\nm} 
& \SI{1064}{\nm} 
& \SI{1064}{\nm} \\
\hline
Beam waist at plane mirror 
& $w_\text{plane}$ 
& \SI{0.92}{\mm} 
& \SI{10.33}{\mm} 
& \SI{8.28}{\mm} \\
\hline
Beam waist at concave mirror 
& $w_\text{conc}$ 
& \SI{1.06}{\mm} 
& \SI{10.34}{\mm} 
& \SI{11.64}{\mm} \\
\hline
Rayleigh range 
& $z_R$ 
& \SI{1.72}{\m} 
& \SI{315}{\m} 
& \SI{203}{\m} \\
\hline
Radius of curvature 
& $R_\text{conc}$ 
& \SI{2}{\m} 
& \SI{5}{\kilo\m} 
& \SI{200}{\m} \\
\hline
Aperture radius 
& $a$ 
& \SI{12.5}{\mm} 
& \SI{17.5}{\cm}
& \SI{17.5}{\cm} \\
\hline
Transmission losses 
& $\delta_\text{trans,1}=\delta_\text{trans,2}$ 
& \SI{100}{ppm} 
& \SI{30}{ppm} 
& \SI{200}{ppm} \\
\hline
Diffraction losses 
& $\delta_{\text{diff}}$ 
& \SI{32}{ppm} 
& \SI{0.03}{ppm} 
& \SI{122.39}{ppm} \\
\hline
Clipping losses 
& $\delta_{\text{clip}}$ 
& \SI{0}{ppm} 
& \SI{0}{ppm} 
& \SI{0}{ppm} \\
\hline
Total round-trip loss 
& $\delta_{\text{total}}$ 
& \SI{232}{ppm} 
& \SI{60.03}{ppm} 
& \SI{522.39}{ppm} \\
\hline
Finesse 
& $\mathcal{F} = {2\pi}/{\delta_{\text{total}}}$ 
& $3 \times 10^4$ 
& $1 \times 10^5$ 
& $1.2 \times 10^4$ \\
\hline
\end{tabular}
\caption{Comparison of the Fabry--Pérot cavity parameters for the WINTER-prototype and the full WINTER setup with the LHC and ALPS-II magnet configurations.}
\label{tab:we_fpc_comparison}
\end{table}

\subsubsection{Thermal Considerations}
\label{appendix:thermal_considerations}

WINTER’s experimental setup employs a high-finesse optical cavity with mirror coatings and substrates engineered to minimize absorption and scattering, enabling stable long-term operation at high intra-cavity power. Under the conditions characterized by the parameters listed in Tab~\ref{tab:we_fpc_comparison}, the system achieves intra-cavity power densities of about \SI[per-mode=symbol]{0.66}{\mega\watt\per\centi\metre\squared}, which lies safely below the upper operational limit of \SI[per-mode=symbol]{1}{\mega\watt\per\centi\metre\squared} demonstrated in the LIGO gravitational-wave detectors under comparable continuous-wave conditions \cite{Billingsley2017}. The mirrors use state-of-the-art dielectric coatings composed of alternating layers of silica (\(\mathrm{SiO_2}\), low refractive index) and titania-doped tantala (\(\mathrm{Ti:Ta_2O_5}\), high refractive index) deposited by ion-beam sputtering, ensuring exceptionally low optical losses. Ultra-pure fused silica substrates with optimized radii of curvature support stable Gaussian modes, while the entire system is housed in high vacuum to suppress convective heat transfer and contamination. Residual heat from absorption in the coatings and substrates is conducted away through the mirror mounts, which provide the primary thermal contact in vacuum and transfer heat to the chamber walls, acting as a heat sink. This integrated approach allows WINTER to maintain high intra-cavity power without thermal distortion or coating degradation.

\end{widetext}

\bibliography{refs}

\begin{thebibliography}{25}%
\makeatletter
\providecommand \@ifxundefined [1]{%
 \@ifx{#1\undefined}
}%
\providecommand \@ifnum [1]{%
 \ifnum #1\expandafter \@firstoftwo
 \else \expandafter \@secondoftwo
 \fi
}%
\providecommand \@ifx [1]{%
 \ifx #1\expandafter \@firstoftwo
 \else \expandafter \@secondoftwo
 \fi
}%
\providecommand \natexlab [1]{#1}%
\providecommand \enquote  [1]{``#1''}%
\providecommand \bibnamefont  [1]{#1}%
\providecommand \bibfnamefont [1]{#1}%
\providecommand \citenamefont [1]{#1}%
\providecommand \href@noop [0]{\@secondoftwo}%
\providecommand \href [0]{\begingroup \@sanitize@url \@href}%
\providecommand \@href[1]{\@@startlink{#1}\@@href}%
\providecommand \@@href[1]{\endgroup#1\@@endlink}%
\providecommand \@sanitize@url [0]{\catcode `\\12\catcode `\$12\catcode `\&12\catcode `\#12\catcode `\^12\catcode `\_12\catcode `\%12\relax}%
\providecommand \@@startlink[1]{}%
\providecommand \@@endlink[0]{}%
\providecommand \url  [0]{\begingroup\@sanitize@url \@url }%
\providecommand \@url [1]{\endgroup\@href {#1}{\urlprefix }}%
\providecommand \urlprefix  [0]{URL }%
\providecommand \Eprint [0]{\href }%
\providecommand \doibase [0]{https://doi.org/}%
\providecommand \selectlanguage [0]{\@gobble}%
\providecommand \bibinfo  [0]{\@secondoftwo}%
\providecommand \bibfield  [0]{\@secondoftwo}%
\providecommand \translation [1]{[#1]}%
\providecommand \BibitemOpen [0]{}%
\providecommand \bibitemStop [0]{}%
\providecommand \bibitemNoStop [0]{.\EOS\space}%
\providecommand \EOS [0]{\spacefactor3000\relax}%
\providecommand \BibitemShut  [1]{\csname bibitem#1\endcsname}%
\let\auto@bib@innerbib\@empty
\bibitem [{\citenamefont {Peccei}\ and\ \citenamefont {Quinn}(1977)}]{peccei_mathrmcp_1977}%
  \BibitemOpen
  \bibfield  {author} {\bibinfo {author} {\bibfnamefont {R.~D.}\ \bibnamefont {Peccei}}\ and\ \bibinfo {author} {\bibfnamefont {H.~R.}\ \bibnamefont {Quinn}},\ }\href {https://doi.org/10.1103/PhysRevLett.38.1440} {\bibfield  {journal} {\bibinfo  {journal} {Physical Review Letters}\ }\textbf {\bibinfo {volume} {38}},\ \bibinfo {pages} {1440} (\bibinfo {year} {1977})}\BibitemShut {NoStop}%
\bibitem [{\citenamefont {Kim}(1979)}]{kim_weak-interaction_1979}%
  \BibitemOpen
  \bibfield  {author} {\bibinfo {author} {\bibfnamefont {J.~E.}\ \bibnamefont {Kim}},\ }\href {https://doi.org/10.1103/PhysRevLett.43.103} {\bibfield  {journal} {\bibinfo  {journal} {Physical Review Letters}\ }\textbf {\bibinfo {volume} {43}},\ \bibinfo {pages} {103} (\bibinfo {year} {1979})}\BibitemShut {NoStop}%
\bibitem [{\citenamefont {Dine}\ \emph {et~al.}(1981)\citenamefont {Dine}, \citenamefont {Fischler},\ and\ \citenamefont {Srednicki}}]{dine_simple_1981}%
  \BibitemOpen
  \bibfield  {author} {\bibinfo {author} {\bibfnamefont {M.}~\bibnamefont {Dine}}, \bibinfo {author} {\bibfnamefont {W.}~\bibnamefont {Fischler}},\ and\ \bibinfo {author} {\bibfnamefont {M.}~\bibnamefont {Srednicki}},\ }\href {https://doi.org/10.1016/0370-2693(81)90590-6} {\bibfield  {journal} {\bibinfo  {journal} {Physics Letters B}\ }\textbf {\bibinfo {volume} {104}},\ \bibinfo {pages} {199} (\bibinfo {year} {1981})}\BibitemShut {NoStop}%
\bibitem [{\citenamefont {Sikivie}(1983)}]{Sikivie_experimental_1983}%
  \BibitemOpen
  \bibfield  {author} {\bibinfo {author} {\bibfnamefont {P.}~\bibnamefont {Sikivie}},\ }\href {https://doi.org/10.1103/PhysRevLett.51.1415} {\bibfield  {journal} {\bibinfo  {journal} {Phys. Rev. Lett.}\ }\textbf {\bibinfo {volume} {51}},\ \bibinfo {pages} {1415} (\bibinfo {year} {1983})}\BibitemShut {NoStop}%
\bibitem [{\citenamefont {Eggemeier}\ \emph {et~al.}(2022)\citenamefont {Eggemeier}, \citenamefont {O'Hare}, \citenamefont {Pierobon}, \citenamefont {Redondo},\ and\ \citenamefont {Wong}}]{eggemeier_minivoids_2022}%
  \BibitemOpen
  \bibfield  {author} {\bibinfo {author} {\bibfnamefont {B.}~\bibnamefont {Eggemeier}}, \bibinfo {author} {\bibfnamefont {C.~A.~J.}\ \bibnamefont {O'Hare}}, \bibinfo {author} {\bibfnamefont {G.}~\bibnamefont {Pierobon}}, \bibinfo {author} {\bibfnamefont {J.}~\bibnamefont {Redondo}},\ and\ \bibinfo {author} {\bibfnamefont {Y.~Y.~Y.}\ \bibnamefont {Wong}},\ }\href@noop {} {\bibinfo {title} {{Axion minivoids and implications for direct detection}}} (\bibinfo {year} {2022}),\ \Eprint {https://arxiv.org/abs/2212.00560} {arXiv:2212.00560 [hep-ph]} \BibitemShut {NoStop}%
\bibitem [{\citenamefont {Vogelsberger}\ and\ \citenamefont {White}(2011)}]{vogelsberger_streams_2011}%
  \BibitemOpen
  \bibfield  {author} {\bibinfo {author} {\bibfnamefont {M.}~\bibnamefont {Vogelsberger}}\ and\ \bibinfo {author} {\bibfnamefont {S.~D.~M.}\ \bibnamefont {White}},\ }\href {https://doi.org/10.1111/j.1365-2966.2011.18224.x} {\bibfield  {journal} {\bibinfo  {journal} {Monthly Notices of the Royal Astronomical Society}\ }\textbf {\bibinfo {volume} {413}},\ \bibinfo {pages} {1419} (\bibinfo {year} {2011})},\ \Eprint {https://arxiv.org/abs/https://academic.oup.com/mnras/article-pdf/413/2/1419/18595473/mnras0413-1419.pdf} {https://academic.oup.com/mnras/article-pdf/413/2/1419/18595473/mnras0413-1419.pdf} \BibitemShut {NoStop}%
\bibitem [{\citenamefont {Mueller}\ \emph {et~al.}(2009)\citenamefont {Mueller}, \citenamefont {Sikivie}, \citenamefont {Tanner},\ and\ \citenamefont {van Bibber}}]{mueller_detailed_2009}%
  \BibitemOpen
  \bibfield  {author} {\bibinfo {author} {\bibfnamefont {G.}~\bibnamefont {Mueller}}, \bibinfo {author} {\bibfnamefont {P.}~\bibnamefont {Sikivie}}, \bibinfo {author} {\bibfnamefont {D.~B.}\ \bibnamefont {Tanner}},\ and\ \bibinfo {author} {\bibfnamefont {K.}~\bibnamefont {van Bibber}},\ }\href {https://doi.org/10.1103/PhysRevD.80.072004} {\bibfield  {journal} {\bibinfo  {journal} {Phys. Rev. D}\ }\textbf {\bibinfo {volume} {80}},\ \bibinfo {pages} {072004} (\bibinfo {year} {2009})}\BibitemShut {NoStop}%
\bibitem [{\citenamefont {Tam}\ and\ \citenamefont {Yang}(2012)}]{tam_production_2012}%
  \BibitemOpen
  \bibfield  {author} {\bibinfo {author} {\bibfnamefont {H.}~\bibnamefont {Tam}}\ and\ \bibinfo {author} {\bibfnamefont {Q.}~\bibnamefont {Yang}},\ }\href {https://doi.org/10.1016/j.physletb.2012.08.050} {\bibfield  {journal} {\bibinfo  {journal} {Physics Letters B}\ }\textbf {\bibinfo {volume} {716}},\ \bibinfo {pages} {435} (\bibinfo {year} {2012})}\BibitemShut {NoStop}%
\bibitem [{\citenamefont {Batllori}\ \emph {et~al.}(2024)\citenamefont {Batllori}, \citenamefont {Gu}, \citenamefont {Horns}, \citenamefont {Maroudas},\ and\ \citenamefont {Ulrichs}}]{wispfi_2024}%
  \BibitemOpen
  \bibfield  {author} {\bibinfo {author} {\bibfnamefont {J.~M.}\ \bibnamefont {Batllori}}, \bibinfo {author} {\bibfnamefont {Y.}~\bibnamefont {Gu}}, \bibinfo {author} {\bibfnamefont {D.}~\bibnamefont {Horns}}, \bibinfo {author} {\bibfnamefont {M.}~\bibnamefont {Maroudas}},\ and\ \bibinfo {author} {\bibfnamefont {J.}~\bibnamefont {Ulrichs}},\ }\href {https://doi.org/10.1103/PhysRevD.109.123001} {\bibfield  {journal} {\bibinfo  {journal} {Phys. Rev. D}\ }\textbf {\bibinfo {volume} {109}},\ \bibinfo {pages} {123001} (\bibinfo {year} {2024})}\BibitemShut {NoStop}%
\bibitem [{\citenamefont {Kimball}\ and\ \citenamefont {Cue}(1986)}]{Primakoff_1935}%
  \BibitemOpen
  \bibfield  {author} {\bibinfo {author} {\bibfnamefont {J.~C.}\ \bibnamefont {Kimball}}\ and\ \bibinfo {author} {\bibfnamefont {N.}~\bibnamefont {Cue}},\ }\href {https://doi.org/10.1103/PhysRevLett.57.1935} {\bibfield  {journal} {\bibinfo  {journal} {Phys. Rev. Lett.}\ }\textbf {\bibinfo {volume} {57}},\ \bibinfo {pages} {1935} (\bibinfo {year} {1986})}\BibitemShut {NoStop}%
\bibitem [{\citenamefont {Borsanyi}\ \emph {et~al.}(2016)\citenamefont {Borsanyi}, \citenamefont {Fodor}, \citenamefont {Guenther}, \citenamefont {Kampert}, \citenamefont {Katz}, \citenamefont {Kawanai}, \citenamefont {Kovacs}, \citenamefont {Mages}, \citenamefont {Pasztor}, \citenamefont {Pittler}, \citenamefont {Redondo}, \citenamefont {Ringwald},\ and\ \citenamefont {Szabo}}]{Borsanyi_2016_mass}%
  \BibitemOpen
  \bibfield  {author} {\bibinfo {author} {\bibfnamefont {S.}~\bibnamefont {Borsanyi}}, \bibinfo {author} {\bibfnamefont {Z.}~\bibnamefont {Fodor}}, \bibinfo {author} {\bibfnamefont {J.}~\bibnamefont {Guenther}}, \bibinfo {author} {\bibfnamefont {K.~H.}\ \bibnamefont {Kampert}}, \bibinfo {author} {\bibfnamefont {S.~D.}\ \bibnamefont {Katz}}, \bibinfo {author} {\bibfnamefont {T.}~\bibnamefont {Kawanai}}, \bibinfo {author} {\bibfnamefont {T.~G.}\ \bibnamefont {Kovacs}}, \bibinfo {author} {\bibfnamefont {S.~W.}\ \bibnamefont {Mages}}, \bibinfo {author} {\bibfnamefont {A.}~\bibnamefont {Pasztor}}, \bibinfo {author} {\bibfnamefont {F.}~\bibnamefont {Pittler}}, \bibinfo {author} {\bibfnamefont {J.}~\bibnamefont {Redondo}}, \bibinfo {author} {\bibfnamefont {A.}~\bibnamefont {Ringwald}},\ and\ \bibinfo {author} {\bibfnamefont {K.~K.}\ \bibnamefont {Szabo}},\ }\href {https://doi.org/10.1038/nature20115} {\bibfield  {journal} {\bibinfo  {journal} {Nature}\ }\textbf {\bibinfo {volume} {539}},\ \bibinfo {pages} {69}
  (\bibinfo {year} {2016})}\BibitemShut {NoStop}%
\bibitem [{\citenamefont {Raffelt}\ and\ \citenamefont {Stodolsky}(1988)}]{raffelt_mixing_1988}%
  \BibitemOpen
  \bibfield  {author} {\bibinfo {author} {\bibfnamefont {G.}~\bibnamefont {Raffelt}}\ and\ \bibinfo {author} {\bibfnamefont {L.}~\bibnamefont {Stodolsky}},\ }\href {https://doi.org/10.1103/PhysRevD.37.1237} {\bibfield  {journal} {\bibinfo  {journal} {Physical Review D}\ }\textbf {\bibinfo {volume} {37}},\ \bibinfo {pages} {1237} (\bibinfo {year} {1988})}\BibitemShut {NoStop}%
\bibitem [{\citenamefont {{V. Anastassopoulos, et al., (CAST Collaboration)}}(2017)}]{CAST_new_2017}%
  \BibitemOpen
  \bibfield  {author} {\bibinfo {author} {\bibnamefont {{V. Anastassopoulos, et al., (CAST Collaboration)}}},\ }\href {https://doi.org/10.1038/nphys4109} {\bibfield  {journal} {\bibinfo  {journal} {Nature Phys.}\ }\textbf {\bibinfo {volume} {13}},\ \bibinfo {pages} {584} (\bibinfo {year} {2017})}\BibitemShut {NoStop}%
\bibitem [{\citenamefont {Drever}\ \emph {et~al.}(1983)\citenamefont {Drever}, \citenamefont {Hall}, \citenamefont {Kowalski}, \citenamefont {Hough}, \citenamefont {Ford}, \citenamefont {Munley},\ and\ \citenamefont {Ward}}]{PDH_technique_1983}%
  \BibitemOpen
  \bibfield  {author} {\bibinfo {author} {\bibfnamefont {R.~W.~P.}\ \bibnamefont {Drever}}, \bibinfo {author} {\bibfnamefont {J.~L.}\ \bibnamefont {Hall}}, \bibinfo {author} {\bibfnamefont {F.~V.}\ \bibnamefont {Kowalski}}, \bibinfo {author} {\bibfnamefont {J.}~\bibnamefont {Hough}}, \bibinfo {author} {\bibfnamefont {G.~M.}\ \bibnamefont {Ford}}, \bibinfo {author} {\bibfnamefont {A.~J.}\ \bibnamefont {Munley}},\ and\ \bibinfo {author} {\bibfnamefont {H.}~\bibnamefont {Ward}},\ }\href {https://doi.org/10.1007/BF00702605} {\bibfield  {journal} {\bibinfo  {journal} {Applied Physics B}\ }\textbf {\bibinfo {volume} {31}},\ \bibinfo {pages} {97} (\bibinfo {year} {1983})}\BibitemShut {NoStop}%
\bibitem [{\citenamefont {{Toptica Photonics AG}}(2024)}]{toptica_private}%
  \BibitemOpen
  \bibfield  {author} {\bibinfo {author} {\bibnamefont {{Toptica Photonics AG}}},\ }\href@noop {} {\bibinfo {title} {Private communication}} (\bibinfo {year} {2024})\BibitemShut {NoStop}%
\bibitem [{\citenamefont {{FEMTO Messtechnik GmbH}}(2024)}]{Femto_FWPR20SI}%
  \BibitemOpen
  \bibfield  {author} {\bibinfo {author} {\bibnamefont {{FEMTO Messtechnik GmbH}}},\ }\href {https://www.femto.de/images/pdf-dokumente/de-fwpr-20-si.pdf} {\emph {\bibinfo {title} {Datasheet: {FWPR-20-SI} Femtowatt Photoreceiver with {Si} Photodiode}}},\ \bibinfo {organization} {FEMTO Messtechnik GmbH},\ \bibinfo {address} {Berlin, Germany} (\bibinfo {year} {2024}),\ \bibinfo {note} {rev. R7/TH, JMa/06MAY2024}\BibitemShut {NoStop}%
\bibitem [{\citenamefont {Bähre}\ \emph {et~al.}(2013)\citenamefont {Bähre}, \citenamefont {Döbrich}, \citenamefont {Dreyling-Eschweiler}, \citenamefont {Ghazaryan}, \citenamefont {Hodajerdi}, \citenamefont {Horns}, \citenamefont {Januschek}, \citenamefont {Knabbe}, \citenamefont {Lindner}, \citenamefont {Notz}, \citenamefont {Ringwald}, \citenamefont {von Seggern}, \citenamefont {Stromhagen}, \citenamefont {Trines},\ and\ \citenamefont {Willke}}]{Bahre2013}%
  \BibitemOpen
  \bibfield  {author} {\bibinfo {author} {\bibfnamefont {R.}~\bibnamefont {Bähre}}, \bibinfo {author} {\bibfnamefont {B.}~\bibnamefont {Döbrich}}, \bibinfo {author} {\bibfnamefont {J.}~\bibnamefont {Dreyling-Eschweiler}}, \bibinfo {author} {\bibfnamefont {S.}~\bibnamefont {Ghazaryan}}, \bibinfo {author} {\bibfnamefont {R.}~\bibnamefont {Hodajerdi}}, \bibinfo {author} {\bibfnamefont {D.}~\bibnamefont {Horns}}, \bibinfo {author} {\bibfnamefont {F.}~\bibnamefont {Januschek}}, \bibinfo {author} {\bibfnamefont {E.~A.}\ \bibnamefont {Knabbe}}, \bibinfo {author} {\bibfnamefont {A.}~\bibnamefont {Lindner}}, \bibinfo {author} {\bibfnamefont {D.}~\bibnamefont {Notz}}, \bibinfo {author} {\bibfnamefont {A.}~\bibnamefont {Ringwald}}, \bibinfo {author} {\bibfnamefont {J.~E.}\ \bibnamefont {von Seggern}}, \bibinfo {author} {\bibfnamefont {R.}~\bibnamefont {Stromhagen}}, \bibinfo {author} {\bibfnamefont {D.}~\bibnamefont {Trines}},\ and\ \bibinfo {author} {\bibfnamefont {B.}~\bibnamefont {Willke}},\ }\href
  {https://doi.org/10.1088/1748-0221/8/09/T09001} {\bibfield  {journal} {\bibinfo  {journal} {Journal of Instrumentation}\ }\textbf {\bibinfo {volume} {8}}\bibinfo  {number} { (09)},\ \bibinfo {pages} {T09001}}\BibitemShut {NoStop}%
\bibitem [{\citenamefont {{Thorlabs, Inc.}}(2022)}]{Thorlabs_DET08CL}%
  \BibitemOpen
\bibfield  {number} {  }\bibfield  {author} {\bibinfo {author} {\bibnamefont {{Thorlabs, Inc.}}},\ }\href {https://www.thorlabs.com/high-speed-free-space-detectors} {\emph {\bibinfo {title} {DET08CL(/M) High-Speed InGaAs Free-Space Photodetector User Guide}}},\ \bibinfo {organization} {Thorlabs, Inc.},\ \bibinfo {address} {Newton, NJ, USA} (\bibinfo {year} {2022}),\ \bibinfo {note} {5 GHz Bandwidth, 800 - 1700 nm}\BibitemShut {NoStop}%
\bibitem [{\citenamefont {Eckhardt}\ and\ \citenamefont {Gerberding}(2022)}]{Gerberding_interferometers_2022}%
  \BibitemOpen
  \bibfield  {author} {\bibinfo {author} {\bibfnamefont {T.}~\bibnamefont {Eckhardt}}\ and\ \bibinfo {author} {\bibfnamefont {O.}~\bibnamefont {Gerberding}},\ }\href {https://doi.org/10.3390/metrology2010007} {\bibfield  {journal} {\bibinfo  {journal} {Metrology}\ }\textbf {\bibinfo {volume} {2}},\ \bibinfo {pages} {98} (\bibinfo {year} {2022})}\BibitemShut {NoStop}%
\bibitem [{\citenamefont {P\~old}\ and\ \citenamefont {Spector}(2020)}]{alps_demonstration_2020}%
  \BibitemOpen
  \bibfield  {author} {\bibinfo {author} {\bibfnamefont {J.~H.}\ \bibnamefont {P\~old}}\ and\ \bibinfo {author} {\bibfnamefont {A.~D.}\ \bibnamefont {Spector}},\ }\href {https://doi.org/10.1140/epjti/s40485-020-0054-8} {\bibfield  {journal} {\bibinfo  {journal} {EPJ Tech. Instrum.}\ }\textbf {\bibinfo {volume} {7}},\ \bibinfo {pages} {1} (\bibinfo {year} {2020})},\ \Eprint {https://arxiv.org/abs/1710.06634} {arXiv:1710.06634 [physics.ins-det]} \BibitemShut {NoStop}%
\bibitem [{\citenamefont {Ejlli}\ \emph {et~al.}(2020)\citenamefont {Ejlli}, \citenamefont {{Della Valle}}, \citenamefont {Gastaldi}, \citenamefont {Messineo}, \citenamefont {Pengo}, \citenamefont {Ruoso},\ and\ \citenamefont {Zavattini}}]{ejlli_pvlas_2020}%
  \BibitemOpen
  \bibfield  {author} {\bibinfo {author} {\bibfnamefont {A.}~\bibnamefont {Ejlli}}, \bibinfo {author} {\bibfnamefont {F.}~\bibnamefont {{Della Valle}}}, \bibinfo {author} {\bibfnamefont {U.}~\bibnamefont {Gastaldi}}, \bibinfo {author} {\bibfnamefont {G.}~\bibnamefont {Messineo}}, \bibinfo {author} {\bibfnamefont {R.}~\bibnamefont {Pengo}}, \bibinfo {author} {\bibfnamefont {G.}~\bibnamefont {Ruoso}},\ and\ \bibinfo {author} {\bibfnamefont {G.}~\bibnamefont {Zavattini}},\ }\href {https://doi.org/https://doi.org/10.1016/j.physrep.2020.06.001} {\bibfield  {journal} {\bibinfo  {journal} {Physics Reports}\ }\textbf {\bibinfo {volume} {871}},\ \bibinfo {pages} {1} (\bibinfo {year} {2020})},\ \bibinfo {note} {the PVLAS experiment: A 25 year effort to measure vacuum magnetic birefringence}\BibitemShut {NoStop}%
\bibitem [{\citenamefont {Altenm\"uller}\ \emph {et~al.}(2024)\citenamefont {Altenm\"uller}, \citenamefont {Anastassopoulos}, \citenamefont {Arguedas-Cuendis}, \citenamefont {Aune}, \citenamefont {Baier}, \citenamefont {Barth}, \citenamefont {Br\"auninger}, \citenamefont {Cantatore}, \citenamefont {Caspers}, \citenamefont {Castel}, \citenamefont {\ifmmode~\mbox{\c{C}}\else \c{C}\fi{}etin}, \citenamefont {Christensen}, \citenamefont {Cogollos}, \citenamefont {Dafni}, \citenamefont {Davenport}, \citenamefont {Decker}, \citenamefont {Desch}, \citenamefont {D\'{\i}ez-Ib\'a\~nez}, \citenamefont {D\"obrich}, \citenamefont {Ferrer-Ribas}, \citenamefont {Fischer}, \citenamefont {Funk}, \citenamefont {Gal\'an}, \citenamefont {Garc\'{\i}a}, \citenamefont {Gardikiotis}, \citenamefont {Giomataris}, \citenamefont {Golm}, \citenamefont {Hailey}, \citenamefont {Hasinoff}, \citenamefont {Hoffmann}, \citenamefont {Irastorza}, \citenamefont {Jacoby}, \citenamefont {Jakobsen}, \citenamefont {Jakov\ifmmode \check{c}\else
  \v{c}\fi{}i\ifmmode~\acute{c}\else \'{c}\fi{}}, \citenamefont {Kaminski}, \citenamefont {Karuza}, \citenamefont {Kostoglou}, \citenamefont {Krieger}, \citenamefont {Laki\ifmmode~\acute{c}\else \'{c}\fi{}}, \citenamefont {Laurent}, \citenamefont {Luz\'on}, \citenamefont {Malbrunot}, \citenamefont {Margalejo}, \citenamefont {Maroudas}, \citenamefont {Miceli}, \citenamefont {Mirallas}, \citenamefont {Navarro}, \citenamefont {Obis}, \citenamefont {\"Ozbey}, \citenamefont {\"Ozbozduman}, \citenamefont {Papaevangelou}, \citenamefont {P\'erez}, \citenamefont {Pivovaroff}, \citenamefont {Rosu}, \citenamefont {Ruiz-Ch\'oliz}, \citenamefont {Ruz}, \citenamefont {Schmidt}, \citenamefont {Schumann}, \citenamefont {Semertzidis}, \citenamefont {Solanki}, \citenamefont {Stewart}, \citenamefont {Vafeiadis}, \citenamefont {Vogel},\ and\ \citenamefont {Zioutas}}]{CAST_Micromegas_2024}%
  \BibitemOpen
  \bibfield  {author} {\bibinfo {author} {\bibfnamefont {K.}~\bibnamefont {Altenm\"uller}}, \bibinfo {author} {\bibfnamefont {V.}~\bibnamefont {Anastassopoulos}}, \bibinfo {author} {\bibfnamefont {S.}~\bibnamefont {Arguedas-Cuendis}}, \bibinfo {author} {\bibfnamefont {S.}~\bibnamefont {Aune}}, \bibinfo {author} {\bibfnamefont {J.}~\bibnamefont {Baier}}, \bibinfo {author} {\bibfnamefont {K.}~\bibnamefont {Barth}}, \bibinfo {author} {\bibfnamefont {H.}~\bibnamefont {Br\"auninger}}, \bibinfo {author} {\bibfnamefont {G.}~\bibnamefont {Cantatore}}, \bibinfo {author} {\bibfnamefont {F.}~\bibnamefont {Caspers}}, \bibinfo {author} {\bibfnamefont {J.~F.}\ \bibnamefont {Castel}}, \bibinfo {author} {\bibfnamefont {S.~A.}\ \bibnamefont {\ifmmode~\mbox{\c{C}}\else \c{C}\fi{}etin}}, \bibinfo {author} {\bibfnamefont {F.}~\bibnamefont {Christensen}}, \bibinfo {author} {\bibfnamefont {C.}~\bibnamefont {Cogollos}}, \bibinfo {author} {\bibfnamefont {T.}~\bibnamefont {Dafni}}, \bibinfo {author} {\bibfnamefont {M.}~\bibnamefont
  {Davenport}}, \bibinfo {author} {\bibfnamefont {T.~A.}\ \bibnamefont {Decker}}, \bibinfo {author} {\bibfnamefont {K.}~\bibnamefont {Desch}}, \bibinfo {author} {\bibfnamefont {D.}~\bibnamefont {D\'{\i}ez-Ib\'a\~nez}}, \bibinfo {author} {\bibfnamefont {B.}~\bibnamefont {D\"obrich}}, \bibinfo {author} {\bibfnamefont {E.}~\bibnamefont {Ferrer-Ribas}}, \bibinfo {author} {\bibfnamefont {H.}~\bibnamefont {Fischer}}, \bibinfo {author} {\bibfnamefont {W.}~\bibnamefont {Funk}}, \bibinfo {author} {\bibfnamefont {J.}~\bibnamefont {Gal\'an}}, \bibinfo {author} {\bibfnamefont {J.~A.}\ \bibnamefont {Garc\'{\i}a}}, \bibinfo {author} {\bibfnamefont {A.}~\bibnamefont {Gardikiotis}}, \bibinfo {author} {\bibfnamefont {I.}~\bibnamefont {Giomataris}}, \bibinfo {author} {\bibfnamefont {J.}~\bibnamefont {Golm}}, \bibinfo {author} {\bibfnamefont {C.~H.}\ \bibnamefont {Hailey}}, \bibinfo {author} {\bibfnamefont {M.~D.}\ \bibnamefont {Hasinoff}}, \bibinfo {author} {\bibfnamefont {D.~H.~H.}\ \bibnamefont {Hoffmann}}, \bibinfo {author}
  {\bibfnamefont {I.~G.}\ \bibnamefont {Irastorza}}, \bibinfo {author} {\bibfnamefont {J.}~\bibnamefont {Jacoby}}, \bibinfo {author} {\bibfnamefont {A.~C.}\ \bibnamefont {Jakobsen}}, \bibinfo {author} {\bibfnamefont {K.}~\bibnamefont {Jakov\ifmmode \check{c}\else \v{c}\fi{}i\ifmmode~\acute{c}\else \'{c}\fi{}}}, \bibinfo {author} {\bibfnamefont {J.}~\bibnamefont {Kaminski}}, \bibinfo {author} {\bibfnamefont {M.}~\bibnamefont {Karuza}}, \bibinfo {author} {\bibfnamefont {S.}~\bibnamefont {Kostoglou}}, \bibinfo {author} {\bibfnamefont {C.}~\bibnamefont {Krieger}}, \bibinfo {author} {\bibfnamefont {B.}~\bibnamefont {Laki\ifmmode~\acute{c}\else \'{c}\fi{}}}, \bibinfo {author} {\bibfnamefont {J.~M.}\ \bibnamefont {Laurent}}, \bibinfo {author} {\bibfnamefont {G.}~\bibnamefont {Luz\'on}}, \bibinfo {author} {\bibfnamefont {C.}~\bibnamefont {Malbrunot}}, \bibinfo {author} {\bibfnamefont {C.}~\bibnamefont {Margalejo}}, \bibinfo {author} {\bibfnamefont {M.}~\bibnamefont {Maroudas}}, \bibinfo {author} {\bibfnamefont
  {L.}~\bibnamefont {Miceli}}, \bibinfo {author} {\bibfnamefont {H.}~\bibnamefont {Mirallas}}, \bibinfo {author} {\bibfnamefont {P.}~\bibnamefont {Navarro}}, \bibinfo {author} {\bibfnamefont {L.}~\bibnamefont {Obis}}, \bibinfo {author} {\bibfnamefont {A.}~\bibnamefont {\"Ozbey}}, \bibinfo {author} {\bibfnamefont {K.}~\bibnamefont {\"Ozbozduman}}, \bibinfo {author} {\bibfnamefont {T.}~\bibnamefont {Papaevangelou}}, \bibinfo {author} {\bibfnamefont {O.}~\bibnamefont {P\'erez}}, \bibinfo {author} {\bibfnamefont {M.~J.}\ \bibnamefont {Pivovaroff}}, \bibinfo {author} {\bibfnamefont {M.}~\bibnamefont {Rosu}}, \bibinfo {author} {\bibfnamefont {E.}~\bibnamefont {Ruiz-Ch\'oliz}}, \bibinfo {author} {\bibfnamefont {J.}~\bibnamefont {Ruz}}, \bibinfo {author} {\bibfnamefont {S.}~\bibnamefont {Schmidt}}, \bibinfo {author} {\bibfnamefont {M.}~\bibnamefont {Schumann}}, \bibinfo {author} {\bibfnamefont {Y.~K.}\ \bibnamefont {Semertzidis}}, \bibinfo {author} {\bibfnamefont {S.~K.}\ \bibnamefont {Solanki}}, \bibinfo {author}
  {\bibfnamefont {L.}~\bibnamefont {Stewart}}, \bibinfo {author} {\bibfnamefont {T.}~\bibnamefont {Vafeiadis}}, \bibinfo {author} {\bibfnamefont {J.~K.}\ \bibnamefont {Vogel}},\ and\ \bibinfo {author} {\bibfnamefont {K.}~\bibnamefont {Zioutas}} (\bibinfo {collaboration} {CAST Collaboration}),\ }\href {https://doi.org/10.1103/PhysRevLett.133.221005} {\bibfield  {journal} {\bibinfo  {journal} {Phys. Rev. Lett.}\ }\textbf {\bibinfo {volume} {133}},\ \bibinfo {pages} {221005} (\bibinfo {year} {2024})}\BibitemShut {NoStop}%
\bibitem [{\citenamefont {Kogelnik}\ and\ \citenamefont {Li}(1966)}]{Kogelnik:66}%
  \BibitemOpen
  \bibfield  {author} {\bibinfo {author} {\bibfnamefont {H.}~\bibnamefont {Kogelnik}}\ and\ \bibinfo {author} {\bibfnamefont {T.}~\bibnamefont {Li}},\ }\href {https://doi.org/10.1364/AO.5.001550} {\bibfield  {journal} {\bibinfo  {journal} {Appl. Opt.}\ }\textbf {\bibinfo {volume} {5}},\ \bibinfo {pages} {1550} (\bibinfo {year} {1966})}\BibitemShut {NoStop}%
\bibitem [{\citenamefont {Fox}\ and\ \citenamefont {Li}(1961)}]{Fox_Li_1961}%
  \BibitemOpen
  \bibfield  {author} {\bibinfo {author} {\bibfnamefont {A.~G.}\ \bibnamefont {Fox}}\ and\ \bibinfo {author} {\bibfnamefont {T.}~\bibnamefont {Li}},\ }\href {https://doi.org/10.1002/j.1538-7305.1961.tb01625.x} {\bibfield  {journal} {\bibinfo  {journal} {The Bell System Technical Journal}\ }\textbf {\bibinfo {volume} {40}},\ \bibinfo {pages} {453} (\bibinfo {year} {1961})}\BibitemShut {NoStop}%
\bibitem [{\citenamefont {Billingsley}\ \emph {et~al.}(2017)\citenamefont {Billingsley}, \citenamefont {Yamamoto},\ and\ \citenamefont {Zhang}}]{Billingsley2017}%
  \BibitemOpen
  \bibfield  {author} {\bibinfo {author} {\bibfnamefont {G.}~\bibnamefont {Billingsley}}, \bibinfo {author} {\bibfnamefont {H.}~\bibnamefont {Yamamoto}},\ and\ \bibinfo {author} {\bibfnamefont {L.}~\bibnamefont {Zhang}},\ }in\ \href {https://dcc.ligo.org/public/0140/P1700029/005/4869-Billingsley-v5.pdf} {\emph {\bibinfo {booktitle} {Proceedings of the American Society for Precision Engineering (ASPE) 2017 Spring Topical Meeting}}}\ (\bibinfo {year} {2017})\ pp.\ \bibinfo {pages} {78--83},\ \bibinfo {note} {lIGO Document P1700029-v5}\BibitemShut {NoStop}%
\end{thebibliography}%

\end{document}